\documentclass{acm_proc_article-sp}

\usepackage{graphicx}
\usepackage{times}
\usepackage[latin1]{inputenc}
\usepackage{color}

\usepackage{subfigure}

\newcommand{\inst}{\mbox{${\it inst\/}$}}  

\newcommand{\nop}[1]{}
\newcommand{\eat}[1]{}

\def\blackbox{{\rule{0.5mm}{0.5mm}}}
\def\qed{\hspace*{\fill}\blackbox}
\newcommand{\condbox}{\qed}

\newcommand{\ra}{\mbox{$\,\rightarrow\,$}}

\newcommand{\lla}{\mbox{$\longleftarrow$}}
\newcommand{\vlla}{\mbox{$\longleftarrow$\hspace*{-0.25ex}---------------}}

\newtheorem{theorem}{Theorem}

\newtheorem{lemmc}{Lemma}
\newtheorem{proposition}{Proposition}

\newtheorem{defn}{Definition}

\newcommand{\crunchbegin}{}
\newcommand{\crunchend}{}   

\newenvironment{namedtheorem}[1]{\begin{theorem}{\bf [#1]\hspace{1mm}:} \begin{rm} \crunchbegin }{ \crunchend \end{rm} \end{theorem}}

\newtheorem{example}{Example}

\newenvironment{defwithname}[1]{\begin{defn} {\bf [#1]} \begin{rm} \crunchbegin}{\condbox \crunchend \end{rm} \end{defn}}

\newenvironment{namedexample}[1]{\begin{example}{\bf [#1]\hspace{1mm}:} \begin{rm} \crunchbegin }{ \condbox \crunchend \end{rm} \end{example}}

\def\blackbox{{\rule{1.5mm}{1.5mm}}}

\newcommand{\comment}[1]{}

\title{HepToX: Heterogeneous Peer to Peer XML Databases~\titlenote{This 
paper was submitted to VLDB 2005. An earlier version was 
submitted to SIGMOD 2005.}}

\author{Angela~Bonifati
\\Icar CNR, Italy
\\e-mail: bonifati@icar.cnr.it
\and Qing~(Elaine)~Chang
\\UBC, Canada
\\e-mail: echang@cs.ubc.ca
\and Terence~Ho
\\UBC, Canada
\\e-mail: terenho@cs.ubc.ca
\and Laks~V.S.~Lakshmanan
\\UBC, Canada
\\e-mail: laks@cs.ubc.ca
}

\begin{document}
\thispagestyle{empty}
\onecolumn
\begin{LARGE}
{\noindent\bf COVER PAGE} \\

{\noindent Technical Report Nr.: \bf UBC TR-2005-15} \\
                                                                                          
{\noindent Paper title: \bf ``Heptox: Heterogeneous Peer to Peer XML Databases''}\\
                                                                                          
{\noindent List of Authors: \bf Angela Bonifati (Icar CNR, Italy),
Qing~(Elaine)~Chang (UBC, Canada), Terence~Ho (UBC, Canada),
and Laks~V.S.~Lakshmanan (UBC, Canada)
}\\

{\noindent First Version Dated: \bf November 2004}\\

{\noindent Revised Version Dated: \bf February 2005}\\

\end{LARGE}

\maketitle

%




\begin{abstract} 
We study a collection of heterogeneous XML databases 
maintaining similar and related information, exchanging data via a 
peer to peer overlay network. In this setting, a 
mediated global schema is unrealistic. Yet, users/applications 
wish to query the databases via one peer using its schema. 
We have recently developed HepToX, a P2P Heterogeneous XML database system. 
A key idea is that whenever a peer enters the system, it establishes an 
acquaintance with a small number of peer databases, possibly with different 
schema. The peer administrator provides correspondences between the local 
schema and the acquaintance schema using an informal and intuitive 
notation of arrows and boxes. We develop a novel algorithm that infers 
a set of precise mapping rules between the schemas from these visual annotations. 
We pin down a semantics of query translation given such mapping rules, and 
present a novel query translation algorithm for a simple but expressive 
fragment of XQuery, that employs the mapping rules in either direction. 
We show the translation algorithm is correct. Finally, we demonstrate the 
utility and scalability of 
our ideas and algorithms with a detailed set of experiments 
on top of the Emulab, a large scale P2P network emulation testbed. 
\end{abstract} 


                                                                                

\vspace*{-1ex} 
\section{Introduction}
\label{sec-intro} 

\begin{figure*}[htb]
\centering
\centerline{\scalebox{0.75}{\includegraphics{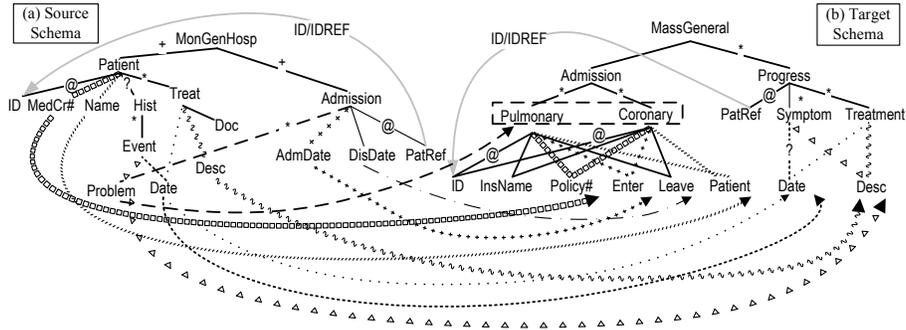}}} 
\caption{Mapping two Heterogeneous Peer DTDs. To minimize clutter,
some arrows between corresponding tags have been omitted (e.g., @ID to @ID); 
every unlabeled edge is labeled `1' by 
default.}\label{fig-running-eg}
\end{figure*}

Consider a large scale data sharing setting where several XML data sources in a similar domain 
(e.g., hospitals/medical centers) 
need  to share (parts of) their data. The source schemas can be quite 
different. 
Figure~\ref{fig-running-eg} shows an 
example of two hospital DTDs expressed as graphs, adapted from \cite{bernstein-etal-webdb02}. 
(See~\cite{med03,peerDB-ng-etal-icde03} 
for similar healthcare examples of P2P DBMS.)  
%
Ignore the cross arrows linking the DTDs for now. 
The first DTD shows how information 
is represented in the {\tt Montreal General Hospital} database. Every patient is assigned a unique 
{\tt ID} by the hospital and has a unique Medicare number ({\tt MedCr\#}). 
Note that symptoms of 
problems experienced by patients and the treatments they are 
administered are all grouped under patients. Data pertaining to patient admissions 
is maintained separately and is captured via the ID/IDREF link {\tt @PatRef}$\ra${\tt @ID} 
(shown as a solid grey arrow in Figure~\ref{fig-running-eg}). 

On the other hand, a patient who is moved 
from Montreal to Boston 
needs to establish early enough 
an acquaintance with the {\tt Mass General Hospital} schema.
Indeed, the patient database at the {\tt Mass General Hospital} in Boston is
organized quite differently.
In this schema, at admission time patients 
are classified on the 
basis of their main complaint (pulmonary, coronary, or other illnesses not shown in 
the figure). Under this classification, the usual patient details such as 
name, hospital ID, insurance number, insurer, admission and discharge dates 
are stored. Progress of patients during their stay in the hospital is recorded: 
patients' history of health problems as well as the treatments administered are 
tracked. All of this information is connected to the patients via the ID/IDREF link 
{\tt @PatRef}$\ra${\tt @ID} (solid grey arrow, Figure~\ref{fig-running-eg}).

Suppose we have several heterogeneous schemas, which describe the 
data of several health/medical data sources.
A natural strategy for data exchange between the sources 
is to embed them in a loosely coupled environment. 
The loose coupling is crucial for allowing maximum flexibility, 
with no hard commitments to conform to any rigid requirements.  
Currently, there is significant interest in peer-to-peer (P2P) database systems 
\cite{miller-sigrec, med03, lenzP2P, millerP2P03, bernstein-etal-webdb02, millerClio03,alon04}. 
A P2P database system is
essentially a collection of autonomous database systems connected by a P2P 
network, with the flexibility that peers may enter or leave the network at will. 
In this paper, we assume that any
peer at the time of entering the P2P network provides a mapping
between its schema and a small number of the existing peer schemas. The peers chosen
by the entering peer are called its {\em acquaintances} \cite{miller-sigrec}. E.g., the {\tt Montreal
General} peer may provide ``correspondences'' between the concepts (elements and
attributes) in its schema and those in the {\tt Mass General} schema.
The next question is who will provide this mapping and how. 
We could make use of (semi-)automatic schema mapping tools \cite{rahm01,potti03,bernstein01}. 
However, one might wish to customize or even override the associations
produced by these tools. To allow for this, we consider a model 
where a peer database
administrator supplies  simple and intuitive correspondences between the peer schema and a 
small number of acquaintance schemas. 
The correspondences are specified
in the form of arrows and boxes, illustrated 
in Figure~\ref{fig-running-eg}, and will be explained in Section~\ref{sec-eg}.
We use the correspondences as a basis for automatically
inferring a {\em mapping expression}, expressed in the form of Datalog-like
rules. The peer DBA may (but doesn't have to)  examine the rules and make any necessary adjustments. 
In this way, joining a P2P DBMS is made a lightweight operation. 

\nop{ 
Note that a P2P database system is significantly different from a PDMS introduced in~\cite{med03,wwwhalevy03}, as joining a PDMS is an heavyweight operation, 
whereas in our framework
all is needed to enter the system is drawing simple schema-to-schema arrows. 
}

There may be considerable differences in the way the peers organize their 
data, including differences in data representations (e.g., stock 
names instead of ticker symbols, different units, etc.), 
as well as differences in underlying schemas 
(e.g., group treatments under the patients receiving them as opposed to maintaining 
separate lists of patients and treatments and linking them with ID/IDREF links). 
%
%
Differences in data representation were addressed by previous work~\cite{millerP2P03}
by introducing the so-called mapping tables. These are 
aimed at mapping value aliases in P2P networks, which are
orthogonal to schema mappings. 
Differences in schemas have also been studied in Clio~\cite{millerClio03} 
and in Piazza~\cite{med03, wwwhalevy03}. 
However,~\cite{millerClio03} focuses on deriving mappings across schemas and 
uses them to translate the data accordingly. Query translation across
the inferred mappings is not 
considered. Furthermore, the kind of 
heterogeneity addressed in \cite{millerClio03} is limited. 
%
%
Piazza~\cite{med03, wwwhalevy03} 
addresses the schema mediation problem within a community of peers, where entering peers  
establish semantic mappings with existing peers, in an XQuery-like
language. 
Our mappings are to be considered in a  large P2P framework. 
For this reason, we 
do not expect a peer DBA/user to know the rule machinery at all, 
whereas it is reasonable to assume s/he can draw arrows and boxes. 
%
%
Piazza is again limited in the extent of schema heterogeneity it can handle. A detailed comparison with this and 
other related work appears in Section~\ref{sec-rw}. 

Our further goal is to permit users and applications of any peer database to access 
data items of interest by simply posing a query to their peer, regardless of the 
location of the data items or the schema under which they are organized. 
In other words, the existence of numerous peers and their schemas should all 
be transparent to the user/application posing the query. E.g., for answering the 
query ``what are the treatments administered to patients admitted with a 
coronary illness?'', we want to manipulate data from all peers ``visible'' to the 
original peer, containing logically relevant information. Clearly, queries posed 
to a peer need to be translated appropriately so as to run on other peers. 

The {\sl major challenges} we address are the following: (1) How do we cope with the 
heterogeneity in the peer schemas in the context of P2P data sharing? 
(2) What is the exact semantics for the evaluation of peer queries, and how do we 
translate queries across peers, given the heterogeneity of their schemas? 
(3) How can we evaluate queries 
correctly and efficiently over the network? 

Summarizing, we make the following contributions. 

\begin{itemize} 
\item We assume correspondences between a pair of peer DTDs are specified 
diagrammatically. We develop a novel algorithm for 
inferring mappings between the DTDs, couched in the form of
Datalog-like rules and discuss the significance of the class of
transformations captured by the rules 
(Section~\ref{sec-arrow2rules}). 

\item We define the semantics of peer queries. We develop a novel query translation 
algorithm that handles a simple but significant fragment of XQuery and show that it is correct 
w.r.t.\ the above semantics. We illustrate our algorithm with examples 
(Section~\ref{sec-query-trans}). Translation is non-trivial even for
the  XQuery fragment considered. 

\item We have developed HepToX,
\footnote{Pronounced Hep Talk: heterogeneous peers talk!} 
a HEterogeneous Peer TO peer Xml database system,  
incorporating the ideas developed in this 
paper. We describe our implementation of HepToX, including the strategies employed for 
efficient query evaluation. We ran an extensive set of 
experiments to measure the effectiveness of our query translation algorithm, as well as the scalability of 
our approach. We discuss the results and the lessons learned (Section~\ref{sec-exper}).  
\end{itemize} 

Section~\ref{sec-eg} gives a motivating example. Related work is discussed in 
Section~\ref{sec-rw}, while Section~\ref{sec-summ} summarizes the paper and 
discusses future research. 

\vspace*{-1ex} 
\section{A Motivating Example} 
\label{sec-eg} 

Revisit the example discussed in the introduction (Figure~\ref{fig-running-eg}). 
The arrows provide informal correspondences between the two hospital 
DTDs, and are illustrated with different types of arrows, for clarity. Henceforth we refer to the 
Montreal General DTD as MonG and the Mass General DTD as MasG. 
The arrows capture simple 1-1 
correspondences between terms such as ``{\tt MedCr\#} in the first DTD to 
{\tt Policy\#} in the second'' and ``{\tt Name} in the first to 
{\tt Patient} in the second''. 
The correspondences can be naturally understood as 
propagating from the leaves up to the their parent/ancestor  elements as in \cite{rahm01}. 
Arrows between leaf elements of tree DTDs have a 
straightforward meaning. Since our DTDs are DAGs, we use different
arrow types  for disambiguation. 
E.g., consider the correspondence between {\tt Desc} in the 
two DTDs. {\tt Desc} has a unique parent in first DTD while it has two 
parents in the second. For disambiguation, we use the same arrow type  for the 
edges {\tt Treat}/{\tt Desc} in MonG, {\tt Treatment}/{\tt Desc} in MasG, 
and the arrow connecting them. Similarly, we use the same arrow type 
for the edges {\tt Event}/{\tt Problem} in MonG, {\tt Symptom}/{\tt Desc} in MasG, and 
the arrow connecting them. Other arrows can be understood similarly. 
E.g., {\tt Event}/{\tt Date} 
and {\tt Treat}/{\tt Date} are matched to their counterparts in the second DTD. 
Finally, consider the (thick dashed) box enclosing the {\tt Pulmonary} and {\tt Coronary} nodes. 
There is a thick dashed arrow matching {\tt Admission}/{\tt Problem} to this box. What it 
says is that `Pulmonary' and `Coronary' correspond to values of the element 
{\tt Admission}/{\tt Problem} in the first database. Again, recall there may be other illnesses 
(corresponding to values of {\tt Admission}/{\tt Problem}) not shown in the figure for brevity. 
In effect, illnesses which may be instances of {\tt Admission}/{\tt Problem} in the first database 
correspond to tags in the second database, but {\sl there is no assumption that the set of 
illnesses occurring in the two databases are the same or even overlap.} 

Boxes are used to group together tags of nodes in a DTD that correspond to 
instances of a tag in another DTD, as part of the correspondence specification. 
Arrows specify a simple one-to-one correspondence between the identified concepts. 
{\sl However, arrows and boxes in and of themselves do {\sf not} tell us how a database 
that conforms to a DTD may be transformed to one that conforms to the other DTD.}  
Why do we care about this transformation any way? The reason is that this transformation 
is closely tied to the semantics of query answering, as we will see in Section~\ref{sec-qt-sem}. 
While we will give an algorithm for inferring mapping expressions from correspondences 
specified using arrows and boxes, here we informally explain the mapping between the 
two DTDs in Figure~\ref{fig-running-eg}. 

Note that in the MonG DTD, patients' history (problems and dates) and the 
treatments they undergo are both nested under patients. All admission information is 
maintained separately and linked to the appropriate patient via the patient's ID. 
In the MasG database, treatment and history information (symptoms) is separated 
out from patients and linked to them via their ID. Additionally, patients are represented 
along with the rest of the admission data, but this data is classified based on the 
type of problem/illness identified at the time of admission. Whatever mapping expressions 
we use should be capable of transforming instances of one of these schemas/DTDs to those of the other. 

\begin{figure}[ht]
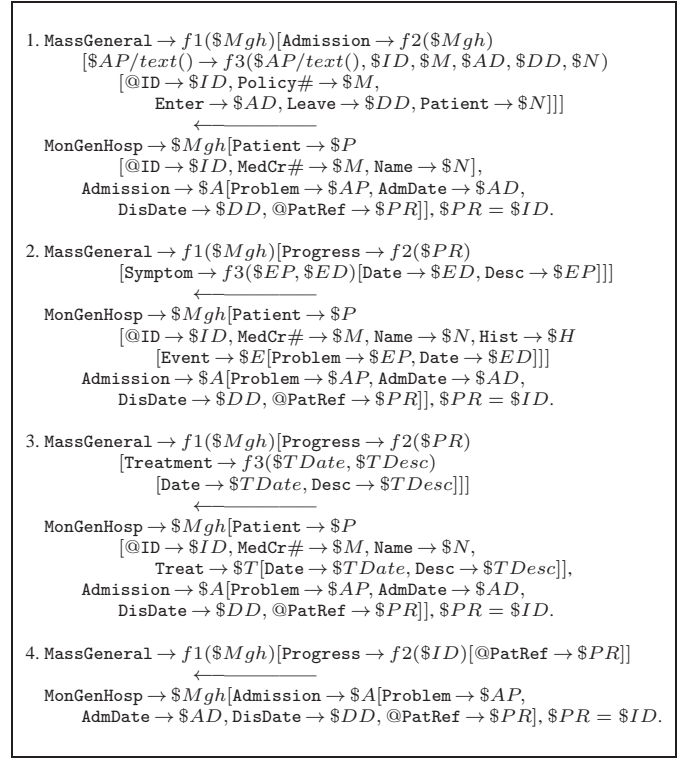
 
\framebox{ 
\begin{minipage}{3.25in} 
\begin{scriptsize}
\begin{tabbing} 
00\=1234\=1234\=1234\=1234\=1234\=1234\=12341234123412341234123412341234 \kill \\ 
1. \>${\tt MassGeneral} \ra f1(\$Mgh)[{\tt Admission} \ra f2(\$Mgh)$\\ 
  \>\>$[\$AP/text() \ra f3(\$AP/text(), \$ID, \$M, \$AD, \$DD, \$N)$ \\ 
    \>\>\>$[{\tt @ID} \ra \$ID, {\tt Policy\#} \ra \$M,$ \\ 
    \>\>\>\> ${\tt Enter} \ra \$AD, {\tt Leave} \ra \$DD, 
                        {\tt Patient} \ra \$N]]]$  \\ 
\>\>\>\>\>$\vlla$ \\
\>${\tt MonGenHosp} \ra \$Mgh[{\tt Patient} \ra \$P$\\ 
      \>\>\>$[{\tt @ID} \ra \$ID, {\tt MedCr\#} \ra \$M, {\tt Name} \ra \$N],$ \\ 
      \>\>${\tt Admission} \ra \$A[{\tt Problem} \ra \$AP, {\tt AdmDate} \ra \$AD, $ \\ 
      \>\>\>${\tt DisDate} \ra \$DD, {\tt @PatRef} \ra \$PR]]$, $\$PR = \$ID$. \\ 
\\ 
2. \>${\tt MassGeneral} \ra f1(\$Mgh)[{\tt Progress} \ra f2(\$PR)$ \\  
    \>\>\>$[{\tt Symptom} \ra f3(\$EP, \$ED)[{\tt Date} \ra \$ED, {\tt Desc} \ra \$EP]]]$ \\ 
\>\>\>\>\>$\vlla$ \\
\>${\tt MonGenHosp} \ra \$Mgh[{\tt Patient} \ra \$P$\\ 
      \>\>\>$[{\tt @ID} \ra \$ID, {\tt MedCr\#} \ra \$M, {\tt Name} \ra \$N, {\tt Hist} \ra \$H$ \\
    \>\>\>\>$[{\tt Event} \ra \$E[{\tt Problem} \ra \$EP, {\tt Date} \ra \$ED]]] $\\
      \>\>${\tt Admission} \ra \$A[{\tt Problem} \ra \$AP, {\tt AdmDate} \ra \$AD, $ \\ 
      \>\>\>${\tt DisDate} \ra \$DD, {\tt @PatRef} \ra \$PR]]$, $\$PR = \$ID$. \\ 
\\ 
3. \>${\tt MassGeneral} \ra f1(\$Mgh)[{\tt Progress} \ra f2(\$PR)$ \\  
    \>\>\>$[{\tt Treatment} \ra f3(\$TDate, \$TDesc)$ \\ 
    \>\>\>\>$[{\tt Date} \ra \$TDate, {\tt Desc} \ra \$TDesc]]]$  \\ 
\>\>\>\>\>$\vlla$ \\
\>${\tt MonGenHosp} \ra \$Mgh[{\tt Patient} \ra \$P$\\ 
      \>\>\>$[{\tt @ID} \ra \$ID, {\tt MedCr\#} \ra \$M, {\tt Name} \ra \$N, $ \\
    \>\>\>\>${\tt Treat} \ra \$T[{\tt Date} \ra \$TDate, {\tt Desc} \ra \$TDesc]]$, \\
      \>\>${\tt Admission} \ra \$A[ {\tt Problem} \ra \$AP, {\tt AdmDate} \ra \$AD,$ \\ 
      \>\>\>${\tt DisDate} \ra \$DD, {\tt @PatRef} \ra \$PR]]$, $\$PR = \$ID$. \\ 
\\ 
4. \>${\tt MassGeneral} \ra f1(\$Mgh)[{\tt Progress} \ra f2(\$ID)[{\tt @PatRef} \ra \$PR]]$ \\  
\>\>\>\>\>$\vlla$ \\ 
\>${\tt MonGenHosp} \ra \$Mgh[{\tt Admission} \ra \$A [{\tt Problem} \ra \$AP,$\\ 
 \>\>${\tt AdmDate} \ra \$AD,{\tt DisDate} \ra \$DD, {\tt @PatRef} \ra \$PR]$, $\$PR = \$ID$. \\ 
\end{tabbing} 
\end{scriptsize}  
\end{minipage} 
} 
\label{fig-mapping-rules} 
\vspace*{-1.5ex}
\caption{Mapping Rules between the schemas in Figure~\ref{fig-running-eg}.} 
\vspace*{-2.25ex}
\end{figure}

Our algorithm, to be given in Section~\ref{sec-arrow2rules}, will produce the 
mapping expression shown in Figure~\ref{fig-mapping-rules}, expressed as Datalog-like rules, 
($\langle$rule head$\rangle$ $\lla$ $\langle$rule body$\rangle$), 
adapted for tree structured data. The counterpart of datalog predicates are 
tree expressions, defined and explained below. 

We explain the mapping expression language next. The rules are made up of {\em atoms} of the form 
${\tt Tag} \ra id$, where {\tt Tag} is a tag or a tag variable 
and $id$ is the id associated with 
a node with this tag. Here, $id$ may be a variable or any term of the form 
$f(\$v_1, ..., \$v_n)$, for some variables $\$v_i$ and some Skolem function 
$f$. 
Atoms can be nested inside other atoms, thus expressing nesting, 
while a comma-separated list of atoms is used for expressing the subelements of a 
given element. Attributes are preceded with a `{\tt @}'. Before we explain the 
meaning of the rules, it's important to bear in mind that the rules and the mapping 
are {\em not} intended for {\sl physically} transforming data from one source's schema to another. 
As pointed out in \cite{miller-sigrec}, they are rather intended for expressing the semantics 
of data exchange --  if data were to be exchanged from source 1 to 2, how would it 
correspond to the schema of source 2. 

Atoms can be nested to form {\em tree expressions}. Tree expressions are 
either atoms ($t \ra i$) or are of the form 
$t \ra i[TE_1, ..., TE_k]$, where $t \ra i$ is an atom and $TE_i$ are tree 
expressions. In Figure~\ref{fig-mapping-rules}, each rule head is a tree expression while 
the rule body is a conjunction of tree expressions and built-in  predicates ($=, >$, etc.). 
Rule 1 says corresponding to the (unique) root of the MonG source, 
there is a (unique) root in MasG. The uniqueness of the latter follows 
from applying the Skolem function $f1$ to the node id associated with the 
former's root, which is unique. Similarly, there is a unique {\tt Admission} node in MasG, and we have 
used again the root of MonG as the argument of the Skolem function $f2$ for capturing 
this uniqueness. The rule body binds the variable $\$AP$ to the problem of a patient 
at admission time. $\$AP/text()$ extracts the text value associated with node 
$\$AP$. This value (which may have a value from the same domain as `Pulmonary' and 
`Coronary'\footnote{The value need {\em not} be one of these.}) is used to form the 
tag of a new node, whose id is $f3(\$AP/text(), \$ID, \$M, \$AD, \$DD, \$N)$, i.e., 
it is a function of the patient's admission time problem ($\$AP/text()$), 
id ($\$ID$), insurance policy (or medicare) number ($\$M$), admission and discharge 
dates ($\$AD, \$DD$), and name ($\$N$). {\sl Note that the arguments of the Skolem function $f3$ are 
exactly the  single-valued subelements of the {\tt Pulmonary} and 
{\tt Coronary} elements in MasG.}\footnote{Optional elements are handled using marked null values.} 
We do not assume any knowledge of keys in this paper. 
\nop{ 
If we know the key of these elements (e.g., patient ID 
$\$ID$), then we can make the node id of these elements 
a function of this key, e.g., $f'(\$ID)$, for some Skolem function $f'$. 
} 
Note that patient id, name, policy number, admission and 
discharge dates are all matched to their counterparts in MasG. 

Rule 2 maps the patient history consisting of {\tt Problem}s and their {\tt Date}s of 
occurrence (nested in MonG through {\tt Hist}/{\tt Event}) to {\tt Symptom}/{\tt Desc} 
and {\tt Symptom}/{\tt Date} in MasG. Note that in MasG, the {\tt Symptom} 
elements are nested inside a {\tt Progress} element, which has as its id a function of the 
patient ID (via {\tt @PatRef}), i.e., $f2(\$PR), \$PR=\$ID$. Thus, there is 
one {\tt Progress} element per patient. Consequently, {\tt Symptom}s are 
grouped by patient ID. The node ID $f3(\$EP, \$ED)$ used for 
{\tt Symptom} elements shows that for each occurrence of a problem for a 
given patient, a separate {\tt Symptom} element is created. 

Rule 3 maps treatment information from MonG to MasG. {\tt Progress} 
elements are created with id $f2(\$PR)$ just as they are in rule 2. Note that the use of 
the node id $f3(\$TDate, \$TDesc)$ for {\tt Treatment} ensures that for every 
treatment on any date administered to a given patient, the corresponding 
{\tt Treatment} element is nested inside the {\tt Progress} element 
associated with the patient. 

Node id's play a key role: for instance, {\tt Progress} elements are created by rules 2 and 
3 independently. Whenever the id of a {\tt Progress} node created by rule 2 matches 
the one created by rule 3, they refer to one and the same node. For instance, suppose 
$`p5'$ is the ID value of a patient. Then the subtree rooted at the {\tt Progress} node 
$f2(p5)$ created 
by rule 2 and the subtree rooted at the {\tt Progress} node  $f2(p5)$ created by rule 3 are both 
glued at the node $f2(p5)$. More generally, whenever subtrees are created by applications 
of the same or different rules, conceptually all these subtrees are glued together at 
nodes having a common node id. This ensures that the pieces ``computed'' by rules 
are correctly glued together. 

Finally, rule 4 maps {\tt @PatRef} attribute in MonG to {\tt @PatRef} attribute 
in MasG, while equating the {\tt @ID}, {\tt @PatRef} attributes in the MasG DTD. 

So far, we explained meaning of mapping rules. A main challenge is their automatic 
derivation from user supplied informal correspondences, and is addressed in 
Section~\ref{sec-arrow2rules}, where we also briefly discuss the
significance of the class of transformations captured by the rules, 
from an algebraic perspective. 
The next challenge is translating queries posed 
against one peer schema so they can be answered from any source. This is dealt with 
in Section~\ref{sec-query-trans}. Finally, the scalability of the framework for a 
real P2P DBMS setting is empirically established in Section~\ref{sec-exper}.




\vspace*{-1ex} 
\section{Inferring Mapping Rules from Arrows} 
\label{sec-arrow2rules} 

In this section, we address the following question. Given a pair of DTDs, represented 
as graphs, and a set of arrows/boxes relating nodes across the graphs, how can we 
automatically infer a set of rules for mapping instances of one DTD into 
instances of the other. First, observe that owing to missing elements, 
we cannot always map a valid instance of one DTD into a valid instance of 
another. E.g., in Figure~\ref{fig-running-eg}, there are no counterparts for the 
nodes {\tt Doc} and {\tt InsName} in the other DTDs. Recall that the purpose of mapping is 
not for physically transforming data across schemas, but rather to capture the correct semantics 
of data exchange. In this section, we develop an automatic mapping rule inference 
algorithm. 
We will use our running example (Figure~\ref{fig-running-eg}) to illustrate the 
algorithm. 

Suppose we wish to infer mapping rules from a DTD $\Delta_1$ (call it source) 
to another $\Delta_2$ (call it target) based on given correspondences. 
The algorithm consists of the following main modules. (1) Determine groups of nodes in the 
two DTDs such that each group intuitively captures some ``unit'' of information. 
(2) For each group, if the group induces a DAG in the original DTD graph, 
convert it into a tree. Then construct a tree expression describing a unit of information 
following the hierarchical structure of the group. (3) For each target group, identify all minimal 
sets of source groups necessary to populate information into the target tree 
expression structure and construct the rules. We explain these steps below. 

\subsection{Detecting Groups} 
First, we need to determine groups of nodes in both DTDs 
to which variables can be bound in a single tree expression in a rule body 
or a rule head. To appreciate this, consider mapping instances of 
Figure~\ref{fig-running-eg}(a) to those of Figure~\ref{fig-running-eg}(b). 
Suppose we write a rule of the form $\langle\mbox{whatever}\rangle \; \lla$\hspace*{-0.25ex}-- 
${\tt MonGenHosp}\ra\$MonG[{\tt Patient}\ra \$P[...],$  
${\tt Admission}\ra \$A[...]]$. Then we create the objects as per the rule head, for 
each {\sl combination} of patient and admission. Thus, the multiplicity of 
the elements in the original database will not be preserved by this rule. 
This problem will be solved if we write mappings for the following 
groups of nodes separately: $\{{\tt MonGenHosp}, {\tt Patient}, {\tt @ID}, 
{\tt MedCr\#}, {\tt Name}\}$ and 
$\{{\tt MonGenHosp}$, ${\tt Admission}$, ${\tt Problem}$, ${\tt AdmDate}$, 
${\tt DisDate}$, ${\tt @Pat\-Ref}\}$. The module for 
determining groups is given in Figure~\ref{fig-group-module}. Our algorithm 
can actually deal with disjunction in DTDs, but we omit disjunction for 
simplicity. We do not consider cyclic DTDs and do not consider order in XML. 
We next explain the group detection algorithm (Figure~\ref{fig-group-module}). 
Figure~\ref{fig-group-example} shows the groups created. 

\begin{figure} 
\framebox{ 
\begin{minipage}{3.25in} 
\begin{scriptsize}
\begin{tabbing} 
00\=1234\=1234\=1234\=1234\=1234\=1234\=12341234123412341234123412341234 \kill \\ 
Input: A DTD $\Delta$. \\ 
Output: A set of groups of nodes from the graph of $\Delta$. \\ 
1. \>Start from the root.  \\ 
       \>If \>there are multiple outgoing edges, mark the root as $P$; \\ 
       \>Else \>follow the single path (no restriction on edge labels) \\ 
            \>\>until it reaches a node that has multiple outgoing edges; \\ 
            \>\>if there is none, return all nodes in $\Delta$ as one group; \\ 
            \>\>else mark the node found as $P$; \\ 
2. \>Start from node $P$; \\ 
   \>Follow each outgoing edge, and stop when we visit an edge \\ 
        \>\>with label $`*'$ or $`+'$; \\       
   \>Mark the node that this edge points to as a ``stop node''; \\  
   \>If there are multiple edges with label $`*'$ or $`+'$ \\ 
        \>\>pointing into nodes in a box \\ 
        \>\>\>(e.g. Pulmonary, Coronary in Figure~\ref{fig-running-eg})\\ 
        \>\>Mark the box as the stop node; \\ 
3. \>If there is at most one stop node found in $\Delta$, \\ 
   \>\>Group the descendants of current $P$ node and all \\ 
   \>\>the nodes on the path from root to $P$ via previously \\ 
   \>\>recursively marked $P$ nodes (if any); \\ 
   \>Else\\ 
   \>\>For each stop node $S$ \{\\ 
     \>\>\>If there are multiple outgoing edges, \\ 
     \>\>\>\>mark this node as $P$; \\ 
     \>\>\>\>Repeat (2) and (3) to find its stop nodes until it \\ 
     \>\>\>\>reaches leaf nodes or find  at most one stop node; \\ 
     \>\>\>Else \\ 
        \>\>\>\>Group the descendents of current stop node $S$, \\ 
        \>\>\>\>and all the nodes on the path from root to $S$ \\ 
        \>\>\>\>via previous recursively marked $P$ nodes; \}\\ 
4. \>Group the rest of the nodes and their ancestor nodes together. \\ 
\end{tabbing} 
\end{scriptsize}  
\end{minipage} 
} 
\caption{Algorithm for Detecting Groups in DTDs.} 
\label{fig-group-module} 
\end{figure} 

Suppose we apply the group detection algorithm on the MonG DTD 
of Figure~\ref{fig-running-eg}(a). Following step 1, we first visit {\tt MonGenHosp} and 
mark it as $P$ (it has multiple outgoing edges). Then, per Step 2, we visit 
{\tt Patient} and (eventually) {\tt Admission}, and mark them as 
stop nodes (since the edge labels are $*$). According to step 2, {\tt Patient} is again 
(recursively) marked as $P$ ($> 1$ outgoing edge) and its associated 
stop nodes {\tt Event} and {\tt Treat} are found, each of which would again 
recursively be marked as $P$. Applying step 3, since there are no stop nodes reachable from 
{\tt Event}, we group all descendants of {\tt Event}, add all nodes 
on the path from {\tt MonGenHosp} (root) to {\tt Event} and create  the 
group $sg_1$ in Figure~\ref{fig-group-example}, 
where we have qualified the shared elements {\tt Problem} and {\tt Date} 
(i.e., {\tt EProblem} and {\tt EDate}) to distinguish them from admission 
problem and admission date. In a similar 
fashion, we will form the groups $sg_2$ 
and $sg_3$. Finally, by step 4, we will form the group $sg_4$, since 
{\tt @ID, MedCr\#, Name} are the left-over attributes and elements for the recursive call on 
{\tt Patient} marked as a $P$ node. We invite the reader to verify that the algorithm will 
form $tg_1$, $tg_2$, $tg_3$ and $tg_4$ (as shown in Figure~\ref{fig-group-example}) on the 
MasG DTD in Figure~\ref{fig-running-eg}(b).

\begin{figure}[h] 
\framebox{ 
\begin{minipage}{3.25in} 
\begin{scriptsize}
\begin{tabbing} 
00\=1234\=1234\=1234\=1234\=1234\=1234\=12341234123412341234123412341234 \kill \\ 
 $sg_1 = \{{\tt MonGenHosp, Patient, Hist, Event, EProblem, EDate}\}$ \\
 $sg_2 = \{{\tt MonGenHosp, Patient, Treat, TDate, TDesc, Doc}\}$ \\
 $sg_3 = \{{\tt MonGenHosp, Admission, Admission\_Problem, AdmDate}$ \\
\>\>${\tt DisDate, @PatRef}\}$\\
 $sg_4 = \{{\tt MonGenHosp, Patient, @ID, MedCr\#, Name}\}$\\
\\
 $tg_1 = \{{\tt MassGeneral, Admission, Pulmonary, Coronary, @ID, InsName,}$\\  
 \>\>${\tt Policy\#, Enter, Leave, Patient}\}$ \\ 
 $tg_2 = \{{\tt MassGeneral, Progress, Symptom,}$ ${\tt SDate, SDesc}\}$ \\  
 $tg_3 = \{{\tt MassGeneral, Progress, Treatment, TDate, TDesc}\}$ \\
 $tg_4 = \{{\tt MassGeneral, Progress, @PatRef}\}$.\\
\\
 {\bf Pairs of Groups Connected by Arrows:}\\
$(sg_3, tg_1), (sg_4, tg_1), (sg_1, tg_2), (sg_2, tg_3), (sg_3, tg_4)$.\\ 
\end{tabbing} 
\end{scriptsize}  
\end{minipage} 
} 
\vspace*{-1.5ex}
\caption{Example of Group Determination for Figure 1.} 
\label{fig-group-example} 
\vspace*{-1.5ex}
\end{figure} 

If we want to write a mapping from one DTD (call it the source) to the other (call it the target),
e.g., from MonG to MasG,
then we consider every target DTD group and write a rule for it by examining those
groups of the source DTD that are connected to it by arrows. In our running example,
the pairs of groups connected by arrows are shown in Figure~\ref{fig-group-example}.

\vspace*{-1ex} 
\subsection{Generating Tree Expressions} 
The next step is to write tree expressions using the groups identified. 
Before doing so, for each group, we examine the subgraph of the original DTD graph induced 
by the nodes in the group (not counting ID$\ra$IDREF(S) edges). {\sl Our group formation 
algorithm always ensures these subgraphs are connected.} If furthermore, the subgraph is a 
tree, we are ready to write the tree expression for that group. If it is a DAG, 
then we replicate each shared node (recursively) as many times as necessary to create a 
tree structure. E.g., consider a standard DTD for books: ${\tt bib} \ra {\tt book*}$, 
${\tt book} \ra {\tt title\ author\ pub}$, ${\tt author} \ra {\tt name}$, 
${\tt pub} \ra {\tt name}$. It is trivial to see that this whole DTD would be 
flagged as a single group by our algorithm. The structure of this DTD graph is a 
DAG. We replicate the {\tt name} node twice -- once for {\tt author} and once 
for {\tt pub} to create a tree structure. An exception is when the DAG structure is the 
result of multiple nodes in a box sharing the same substructure. E.g., in the target DTD MasG, 
Pulmonary, Coronary, etc., all share the same substructure and they are in a box. 
In this case, we will {\em not} replicate the shared elements. Instead, the use of a 
tag variable will deal with this correctly. 

For each source group, the tree expression 
is written by essentially following the recursive tree structure, using $[...]$ 
to capture the nesting. For every node in the group, we write the expression 
${\tt Tag} \ra \$var$, where {\tt Tag} is the tag name of the node and $\$var$  is a 
new variable. If the node is inside a box (e.g., like Pulmonary in the MasG DTD), then 
we use a tag variable and write $\$Tag \ra \$var$. Group $sg_2$  induces a tree so 
we can immediately write its tree expression as 
${\tt MonGenHosp} \ra \$Mgh$$[{\tt Patient} \ra \$P$$[{\tt Treat} \ra \$T$ \\
$[{\tt Date} \ra \$TDate,{\tt Desc}\ra \$TDesc,$ ${\tt Doc}\ra\$Doc]]]$. 
All the source groups happen to 
induce trees. 

For target groups, we follow the same procedure, replicating nodes if necessary to 
convert any DAGs induced by a group into a tree. Once we have trees, we write 
tree expressions with them. However, there is a major difference w.r.t. source groups, i.e., 
we do {\em not} know the 
node id's to be used in the generated tree expressions. Instead, we 
write the tree expression as a skeleton, leaving the node id's  as `??' for now. As an example, 
the tree expression for $tg_1$ would be ${\tt MassGeneral} \ra ??$ $[{\tt Admission}\ra??$ $[\$Tag \ra 
??$ $[{\tt @ID}\ra ??$ $, ..., $ ${\tt Patient}\ra??]]]$. 
The `??' will be filled in when we write the mapping rules. 
We drop a leaf node from 
consideration if there is no counterpart in the source schema. E.g., {\tt Doc} is one 
such node. 
The module for creating trees and for writing tree expressions is straightforward and is 
not shown. 

\vspace*{-1ex} 
\subsection{Generating Mapping Rules} 
The last step is writing the mapping rules. 
We present the rule construction algorithm in Figure~\ref{fig-form-rules}.

\begin{figure} 
\framebox{ 
\begin{minipage}{3.25in} 
\begin{scriptsize}
\begin{tabbing} 
00\=123\=123\=123\=123\=123\=123\=123123123123123123123123 \kill \\ 
Input: source groups, and target groups \\ 
Output: a set of mapping rules \\ 
For every target group $tg$  \\ 
   \>Let $\{sg_i, ..., sg_j\}$ be the set of source groups connected to\\
   \> $tg$ by arrows\\ 
   \>Let $TE(tg)$ be the tree expression for group $tg$\\ 
   \>1.Start with the rule skeleton \\
      \>\>\>$TE(tg) \lla TE(sg_i), ..., TE(sg_j)$ \\ 
   \>\>Fill in the variables corresponding to leaf positions in $TE(tg)$ \\ 
   \>\>based on the arrows incident on the leaf elements of $tg$ \\ 
   \>2.For root and each of its descendants via only single-valued edges\\
      \>\>Assign their ids as distinct Skolem functions of the root \\
      \>\> variable in the source \\ 
   \>3.For each internal node $inode$ \\        
      \>\>Assign its id as a distinct Skolem function of the variables\\
      \>\>associated with all its single-valued children \\  
      \>\>If any of its single-valued children $sc$ does not belong to $tg$\\
        \>\>\>Trace the source group $sg_p$ with an arrow pointing to $sc$\\
        \>\>\>Add $TE(sg_p)$ in the rule body: \\
          \>\>\>\>$TE(tg) \lla TE(sg_i), ..., TE(sg_j), TE(sg_p)$ \\ 
        \>\>\>Let $scref$ denote the corresponding element of $sc$ in the \\
        \>\>\> source schema \\
        \>\>\>If $scref$ points to a node in the source schema $scid$ and the\\
        \>\>\>source group $sg_q$ that $scid$ belongs to is not in $\{sg_i, ..., sg_j\}$\\ 
          \>\>\>\>Add $TE(sg_q)$ in the rule body: \\
            \>\>\>\>\>$TE(tg) \lla TE(sg_i), ..., TE(sg_j), TE(sg_p), TE(sg_q)$ \\  
        \>\>\>\>Equate the variable $scref$ binds to ($\$scref$) with the \\
        \>\>\>\> variable $scid$ binds to ($\$scid$): $\$scref$=$\$scid/text()$ \\
        \>\>\>\>(if $scid$ is an attribute, then use $\$scref$=$\$scid$ instead)\\
          \>\>\>\>Add $\$scref$ as an argument to the Skolem function \\ 
\>\>\>\> \> on the RHS of $inode$\\
\end{tabbing} 
\end{scriptsize}  
\end{minipage} 
} 
\vspace*{-1.5ex}
\caption{Rule Construction Algorithm.} 
\label{fig-form-rules} 
\end{figure} 

Consider each target group $tg$. Let $\{sg_i, ..., sg_j\}$ be the set of 
source groups connected by arrows to $tg$. Let $TE(g)$ denote the tree 
expression for group $g$. Applying step 1, we start with the rule skeleton $TE(tg) \lla 
TE(sg_i), ..., TE(sg_j)$. Based on the arrows incident on the leaf elements of 
$tg$, we fill in the variables corresponding to leaf positions in $TE(tg)$, i.e., 
the right-hand-side of atoms corresponding to leaf nodes. E.g., for $tg_2$, we 
would start with $TE(tg_2) \,\lla\, TE(sg_1)$. 
The rule body only contains $TE(sg_1)$ since that is the only source group connected to $tg_2$ 
by arrows (see Figure~\ref{fig-group-example}). Based on the arrows, we can fill in the 
right-hand-sides of {\tt Date} and {\tt Desc} in the rule head as $\$ED$ and $\$EP$, respectively: 
${\tt MassGeneral}\ra??$ $[{\tt Progress}\ra??$ $[{\tt Symptom}\ra ??$ $[{\tt Date}\ra??,$ 
${\tt Desc}\ra??]]]$ 
$\lla$\hspace*{-0.25ex}--  
${\tt MonGenHosp}\ra\$M$ $[{\tt Patient}\ra\$P$ $[{\tt Hist}\ra\$H$ 
$[{\tt Event}\ra\$E$ $[{\tt EProblem}\ra\$EP,$ 
${\tt EDate}\ra\$ED]]]]$. 

Next, according to step 2, for root {\tt MassGeneral} and its single-valued child 
{\tt Admission}, we assign
their ids as distinct Skolem functions of the root variable in the source ($\$Mgh$). 

Then we apply step 3: for each 
internal node, we assign its id as a distinct Skolem function of the variables associated 
with all its single-valued children. E.g., for the RHS of {\tt Symptom}, we would use the Skolem 
function $f3(\$EP, \$ED)$. For {\tt Progress}, its (only) single-valued 
child is {\tt @PatRef}, which does not belong to $tg_2$. By following the arrow incident on {\tt @PatRef} we 
trace source group $sg_3$, so we introduce $TE(sg_3)$ in the rule body. Now, {\tt @PatRef} 
in the source schema points to the ID attribute {\tt @ID} of {\tt Patient}, an attribute that 
does not belong to $sg_1$ or $sg_3$. However, {\tt @ID} belongs to $sg_4$ so we also add 
$TE(sg_4)$ to the above rule body. We equate the variable associated with the {\tt @ID} 
attribute in $TE(sg_4)$ with the variable associated with {\tt @PatRef} in $sg_1$. At this point, the rule looks as follows: 

\vspace*{-1.5ex}
\begin{scriptsize}
\begin{tabbing} 
${\tt MassGeneral}\ra f1(\$Mgh)$ $[{\tt Progress}\ra ??$ \\
\hspace{5mm} $[{\tt Symptom}\ra f3(\$ED, \$EP)$ \\ 
\hspace{5mm} $[{\tt Date}\ra\$ED,$ ${\tt Desc}\ra\$EP]]]$ \\

\hspace{14mm} $\vlla$ \\

${\tt MonGenHosp}\ra\$M$ $[{\tt Patient}\ra\$P$ $[{\tt Hist}\ra\$H$ \\ 
\hspace{5mm} $[{\tt Event}\ra\$E$ $[{\tt Problem}\ra\$EP,$ \\
\hspace{5mm} ${\tt Date}\ra\$ED]]]]$,\\ 
${\tt MonGenHosp}\ra\$M$ $[{\tt Admission}\ra\$A$ \\
\hspace{5mm} $[{\tt Problem}\ra\$AP,$ ${\tt AdmDate}\ra\$AD,$  \\
\hspace{5mm} ${\tt DisDate}\ra\$DD,$ ${\tt @PatRef}\ra\$PR]]$, \\
${\tt MonGenHosp}\ra\$M$ $[{\tt Patient}\ra\$P$ \\
\hspace{5mm} $[{\tt @ID}\ra\$I,$ ${\tt Name}\ra\$N,$ \\
\hspace{5mm} ${\tt MedCr\#}\ra\$MC]]$, $\$PR = \$I$. \\
\vspace{-3mm}
\end{tabbing}
\end{scriptsize}
\vspace{-6mm}
Now, for the RHS of {\tt Progress}, we can assign the Skolem function 
$f2(\$PR)$. 
We then refine the rule by identifying nodes/paths shared between two or more tree 
expressions in the body. This would yield rule 2 in Figure 2. Note that the generated 
rules are always {\em safe} -- all variables in the head appear in the body. 
{\sl The main 
contribution of this section is the automatic inference of mapping rules 
that transform tree database instances of one schema into those of another.}  

We next briefly address the question, what is significant or
fundamental about the class of transformations that are captured by
the rules? For lack of space, we merely sketch this idea and refer the
reader to \verb+ftp://ftp.cs.ubc.ca/~laks+ \verb+/algebraicTransformations4Heptox.pdf+ for
more details. The key idea is that the rules capture a class of
database tree transformations that are expressible using the operators
unnest/nest (similar to those for nested relations), flip/flop (which
basically change nesting orders in the schema), and merge/split (which
have a flavor of grouping and ``ungrouping'' a set of nodes. It can be
shown that the rules capture precisely the class of transformations
expressible using these operators together with a few additional
operators like node addition/deletion and tag modification, added for
completeness purposes. 

A comment about optional single valued elements which may form
arguments of Skolem functions. We can model their optionality using
distinct marked null values. The details will appear in the full
paper. 



\vspace*{-1ex} 
\section{Query Translation} 
\label{sec-query-trans} 

In this section, we address two questions. (1) Suppose we have a pair of peer 
XML database sources $p_i$, with DTDs $\Delta_i$ and underlying database instances 
$D_i$, $i = 1, 2$. Suppose we issue a query $Q$ against the DTD of $p_1$ ($p_2$). What does it 
mean for $Q$ to be answered using the database of $p_2$ ($p_1$)? 
(2) Can we translate the query $Q$ against the other peer's DTD (say $\Delta_2$) such that 
the translated query $Q'$ against $\Delta_2$ evaluated on the database at $p_2$ will 
yield the correct answers w.r.t. the semantics captured by the answer to question (1)? 
We also wish to do the translation efficiently. 

\subsection{Query Translation Semantics} 
\label{sec-qt-sem} 

Suppose we have the peers, DTDs, and database instances as defined above. Suppose we 
have mapping rules $\mu$ mapping database instances of $\Delta_1$ to those of 
$\Delta_2$, i.e., $\mu: \Delta_1 \ra \Delta_2$. 
Now, a query $Q$ can be posed against $\Delta_1$ or against $\Delta_2$. 
The following definition captures the correct semantics of query translation for 
these two cases. Let $\inst(\Delta)$ denote the set of  
database instances of $\Delta$. 
\vspace*{-3mm}
\begin{defwithname}{Semantics} 
\label{def-sem} 
{\em Suppose $Q_i$ is a query posed against $\Delta_i$, $i = 1, 2$. Let 
$Q_i^t$ denote a translation of $Q_i$ against $\Delta_j$, $j \neq i$. 
Then $Q_2^t$ is {\em correct} provided $\forall D_1 \in \inst(\Delta_1): \; 
Q_2^t(D_1) = Q_2(\mu(D_1))$. 
The translation $Q_1^t$ is {\em correct} provided $\forall D_2 \in \inst(\Delta_2): \; 
Q_1^t(D_2) = \bigcap_{D_1^k: \mu(D_1^k) = D_2}  \, Q_1(D_1^k))$. 
} 
\end{defwithname}
\vspace*{-3mm}
The translation $Q_2^t$ is correct provided evaluating 
$Q_2$ on the transformed instance $\mu(D_1)$ and evaluating $Q_2^t$ on $D_1$ both yield 
the same results, for all $D_1 \in \inst(\Delta_1)$. Note that in this case, the 
direction of translation is {\sl against} that of the mapping $\mu$. 
This is the easy direction. 
Consider translating a query $Q_1$ posed against $\Delta_1$ to the DTD $\Delta_2$ of 
peer $p_2$. This is {\sl aligned} with the direction of the mapping $\mu$. 
{\sl The complication here is the mapping $\mu$ that transforms instances of $\Delta_1$ 
to those of $\Delta_2$ may not be invertible}. In this case, we require that 
no matter what instance $D_1^k \in \inst(\Delta_1)$ was transformed by $\mu$ to 
$D_2$ (i.e., $\mu(D_1^k) = D_2$) the answer to $Q_1^t$ on $D_2$ should coincide with the 
answer to $Q_1$ on $D_1^k$. This is captured by taking the intersection of the 
latter answers over all possible preimages $D_1^k$ for which $D_2 = \mu(D_1^k)$. 

Note that owing to schema discrepancies between $\Delta_1$ and $\Delta_2$, 
the output of $Q_2(\mu(D_1))$ would ``conform'' to the schema $\Delta_2$. 
What can we say about $Q_2^t(D_1)$? Note that XQuery permits restructuring of output 
so this is not an issue. A similar comment applies to the translation $Q_1^t$ of 
$Q_1$. 
{\sl One of the main contributions of this paper is that our query translation algorithm, 
given in the next section, is correct in the sense defined above, regardless of 
the direction of the mapping rules.}  

\subsection{Query Translation Algorithm} 
\label{sec-qta} 

\nop{Given a mapping $\mu$ from DTD $\Delta_1$ to DTD $\Delta_2$ 
($\mu: \Delta_1 \ra \Delta_2$), 
we give a query translation algorithm that seamlessly 
works when the input query $Q_i$ is posed  
against either $\Delta_1$ or $\Delta_2$. 
The XQuery fragment we consider here is negation-free,
disjunction-free, function-free (including no aggregate functions) XQuery.}

{\bf XQuery fragment considered}: 
{\sl 
The fragment of XQuery we consider corresponds to queries expressible as 
joins of tree patterns (TP)~\cite{sihem01}, where the return arguments correspond 
to leaf nodes of the database. We note that even for this 
simple fragment of XQuery, query translation is far from trivial. We briefly review tree 
patterns next. }

\nop{ 
For convenience in processing XQuery statements, we need a concise 
representation. In this paper, we make use of tree patterns (TP)~\cite{sihem01} 
introduced in the literature as a tool 
for representing the XQuery fragment considered and for translating it. We 
briefly review this notion below. 
} 

A tree pattern is a rooted tree with child and descendant edges, and with nodes 
labeled by variables. Each node variable may be additionally constrained to 
have a specific tag (e.g., name(\$x) = book) and (in case of leaf elements) 
possibly constrained on its value (e.g., \$y/text() = ``123'' or 
\$x/text() $>$ \$y/text()). Figure~\ref{TP}(b) shows an example in which 
some constraints of both types are specified. 
Intuitively, the semantics of a TP is that its nodes are matched 
to the data nodes in an input XML database while preserving tags, 
edge relationships, and any constraints on node contents. Each match 
leads to an answer that consists of the data node that matches a 
specially marked {\em distinguished node} of the TP. E.g., in Figure~\ref{TP}(b) 
the answer should contain {\tt Name} nodes which corresponds to the output
node marked with a dashed box. As a convention, if no explicit distinguished node 
is marked in an example, we intend that all leaf nodes of the TP are distinguished. 


\subsubsection{Translating Tree Patterns} 
Let us now return to query translation. 
Naturally, the details of the translation vary a little depending on whether 
we translate in the direction of the mapping rules or against it. 
Let us first consider translating in the direction of the 
mapping rules (i.e., from the body of each rule to its head). So, we want to 
translate a query $Q_1$ on $\Delta_1$ to one on $\Delta_2$. 
As an overview, we represent a query as (a join of) one or more 
tree patterns and translate 
them one by one. The translation consists of three phases: 
(1) An {\em expansion} phase, where we expand the TP by adding more nodes and 
edges so it can be matched to a rule body. It may also involve ``unwinding'' a 
descendant edge to spell out the intermediate nodes on a path. Essentially, the 
rule body (i.e., the tree expression in the body) is the expanded TP. 
Expansion requires finding a 
substitution that relates variables in the rule body to those in the TP. 
(2) A {\em translation} phase, where we ``apply'' the rule and generate the 
instance of the rule head corresponding to the substitution used. 
(3) A {\em contraction} (or shrinking) phase, where certain nodes are identified 
as redundant and are eliminated from the translated TP. 
We explain these steps in detail below. 

\begin{figure}[htb]
\centering
\includegraphics[height=0.75in,width=3in]{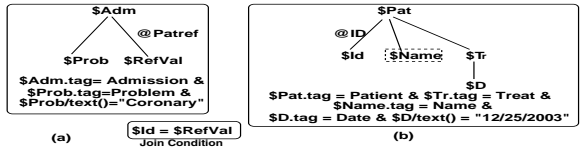}
\caption{Join of two Tree Patterns.}\label{TP}
\end{figure}

\begin{figure}[htb]
\centering
\includegraphics[height=2in,width=3.0in]{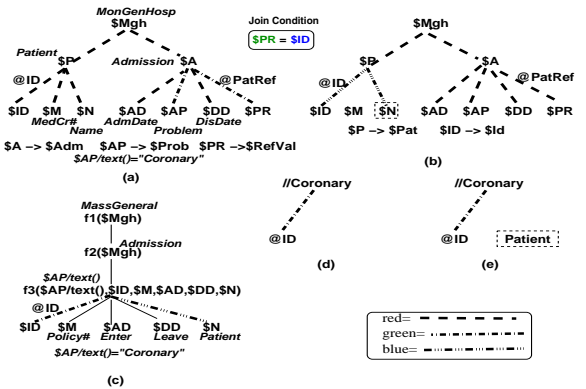}
\caption{Translation of Join of TPs from Figure~\ref{TP}.}\label{subst1}
\end{figure}

\noindent 
{\bf Expansion}: 
Let $t$ be a TP and $r: h_r \lla b_r$ be a mapping rule. Recall that 
the body $b_r$ consists of tree expressions possibly together with some 
comparison predicates (see Figure~\ref{fig-mapping-rules}, e.g.). 
We first find a matching of $t$ to 
$b_r$, which is a substitution that maps variables in $b_r$ to those 
in $t$. This substitution (mapping) may be partial in that $b_r$ may contain 
components which have no counterpart in $t$. E.g., consider Figure~\ref{TP}(a). 
Ignore the join condition for now. 
Let us try to match this TP to the body of rule $1$ in Figure~\ref{fig-mapping-rules}. 
The corresponding expansion is shown in Figure~\ref{subst1}(a). Note that the 
expanded TP is just the body of rule $1$. The part of the expanded TP that was 
originally present in the TP (before expansion) is shown in green while the rest 
(i.e., the new edges coming from expansion) is shown in red (colors are 
represented with line styles as in the legenda). We refer to nodes that were not 
part of the original TP as {\em dummy} nodes. In Figure~\ref{subst1}(a), nodes 
with no incident green edges are precisely the dummy nodes (e.g., \$Mgh, \$P, \$DD, etc.). 

\noindent 
{\bf Translation}: 
Next, we translate the expanded TP by applying the rule to it. The correspondence between 
the variables in the original TP and those in the expanded TP (i.e., the rule body) 
is kept track of by means of a substitution between the two. Figure~\ref{subst1}(a) 
shows the substitution for the TP in question as $\{\$A \ra \$Adm, \$AP \ra \$Prob, 
\$PR \ra \$RefVal\}$. Using this substitution, original query constraints are 
propagated through the translation. In our example, the translated query resulting 
from applying rule $1$ to the expanded TP above is shown in Figure~\ref{subst1}(c) 
(ignore the blue edge for now). For readability, all tag constraints are shown 
concisely by writing the tags next to the appropriate nodes. Note how the 
constraint $\$Prob/text() = ``Coronary''$ is propagated via the substitution 
as $\$AP/text() = ``Coronary''$. Additionally, the condition $\$PR = \$ID$ in the 
body of rule $1$ is used to infer that the attribute child $@ID$ of the node $\$AP/text()$ 
in Figure~\ref{subst1}(c) corresponds to the attribute child $@PatRef$ of the node $\$A$ in 
Figure~\ref{subst1}(a). At this point, the translated query contains Skolem functions at 
certain nodes. By keeping track of the correspondences between the dummy nodes 
in the expanded TP (Figure~\ref{subst1}(a)) and the translated query, we 
can identify nodes that can be eliminated from this query. Specifically, in the 
expanded TP, only nodes $\$AP, \$PR$, and \$$A$ were non-dummy nodes. The 
corresponding nodes in the translated query are $f3(\$AP/text(), \$ID, \$M, \$AD, \$DD, \$N)$ 
(a Skolem function whose arguments include $\$AP/text()$),  
and $\$ID$ (which corresponds to $\$PR$).  Note that there is no counterpart for $\$A$ in 
Figure~\ref{subst1}(c). So, all other nodes in the translated query are dummy nodes and 
can be dropped. More precisely, if a translated query node corresponds to a non-dummy node from the 
expanded TP or is a Skolem function one of whose arguments corresponds to non-dummy node, 
then it is non-dummy. Otherwise, it is 
dummy. 

\noindent 
{\bf Contraction}: 
Having identified dummy nodes in the translated query, we try to drop them. For leaf dummy nodes, 
this is trivial and they are always dropped, while preserving query equivalence. 
An internal dummy node can be dropped provided dropping it and connecting its parent and 
children with descendant axis (`//') will not change the 
corresponding set of {\sl DTD paths}  
connecting them. So, whenever an internal dummy node {\em is} dropped, its parent and 
children are connected by a descendant edge. In Figure~\ref{subst1}(c), all nodes with 
no incident green edges are dummy and can be dropped. Note that the set of paths connecting 
the remaining nodes remains unchanged according to the DTD $\Delta_2$. 
Each remaining (non-dummy) node is 
replaced with its corresponding tag. If the tag is an unconstrained  variable, it is replaced by the 
wildcard `*'. If it is a constrained variable, 
it is replaced by each valid tag satisfying the constraint. Whenever internal dummy nodes 
are dropped, the corresponding child edges are replaced by descendant edges. 
Doing this to the translated query in Figure~\ref{subst1}(c) yields the final contracted 
(shrunk) translated query in Figure~\ref{subst1}(d). So, 
at this point, the final query is {\small \tt //Coronary/@ID}. 

It is easy to verify that when query translation is applied to the TP 
in Figure~\ref{TP}(b), using the body of rule $1$ in Figure~\ref{fig-mapping-rules}, 
we will get the expanded TP in Figure~\ref{subst1}(b), the translated query in 
Figure~\ref{subst1}(c) (with both green and blue edges considered), and 
finally the contracted translated query in Figure~\ref{subst1}(e) (with both 
green and blue edges considered). 

\subsubsection{Translating Joins of Tree Patterns} 
Now, consider the query in Figure~\ref{subst1}(a)-(b), {\sl including} the join 
condition $\$PR = \$ID$. As illustrated above, Figure~\ref{subst1}(a) leads to the 
contracted translated query {\small \tt //Coronary/@ID} while Figure~\ref{subst1}(b) leads 
to the contracted translated query {\small \tt //Coronary[@ID]/Name}.
From the join condition, we deduce that the {\tt @ID} node 
in both queries denote the same value and hence their parent {\small \tt Coronary} nodes must be 
identical. Based on this, we ``stitch'' the two translated queries together by 
identifying their {\small \tt Coronary} nodes and their {\small \tt @ID} nodes. 
So, the final contracted translated TP is isomorphic to the one shown in 
Figure~\ref{subst1}(e) (with both green and blue edges considered). 

\subsubsection{An XQuery Example} 
Let us illustrate how our translation algorithm developed 
so far can handle a simple fragment of XQuery. {\sl This example 
illustrates translation in the direction of the mapping 
rules.}  

\begin{figure}[htb]
   \includegraphics[height=1.5in,width=3in]{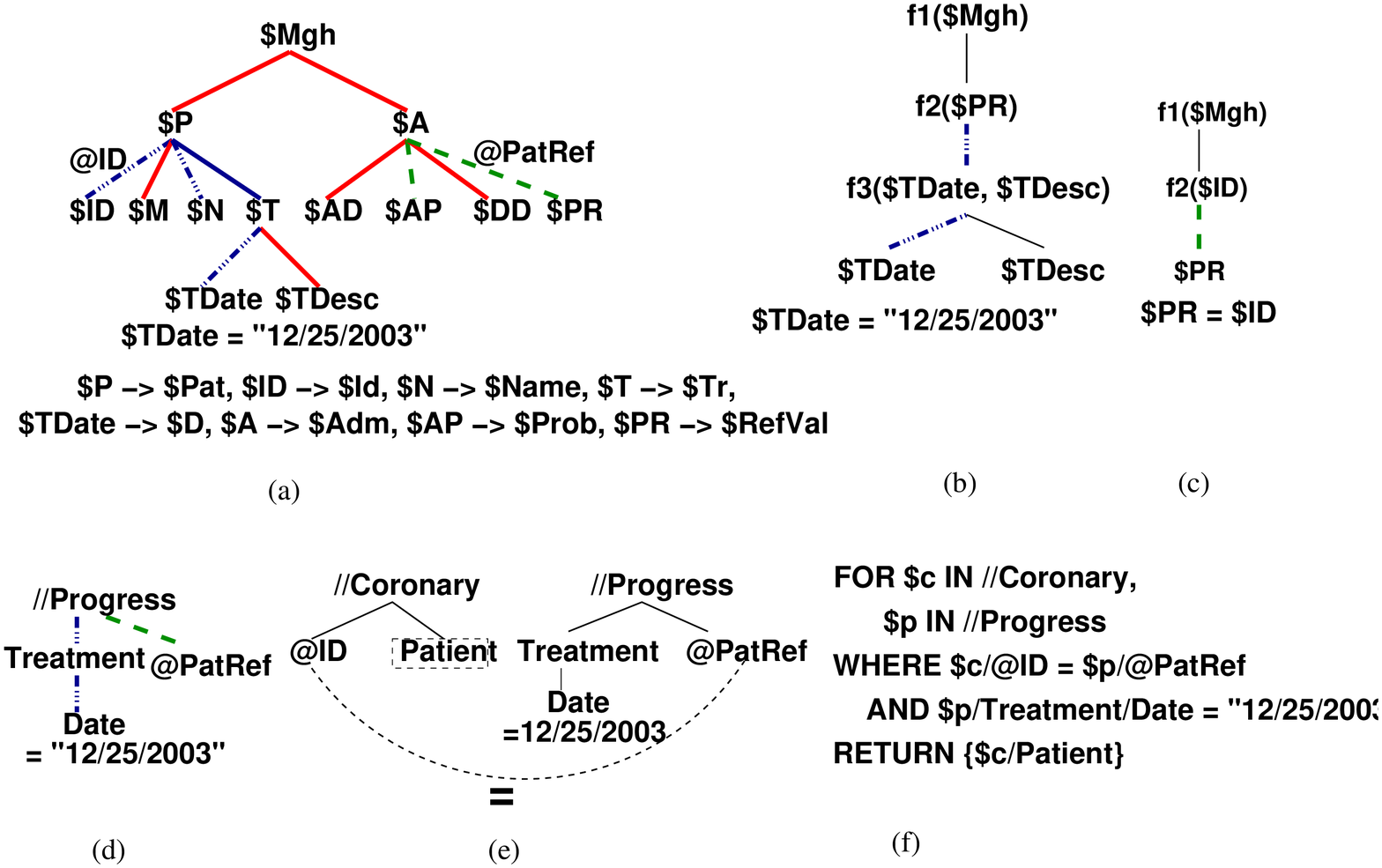}
\vspace*{-1.5ex} 
  \caption{Translation of $Q1$ completed: tag constraints suppressed
    for avoiding clutter; red = solid, thick; green = dashed, thick;
    blue = dashed \& dotted, thick.}
  \label{subst2}
  \vspace{-0.25cm}
  \end{figure}

\begin{namedexample}{XQuery Forward} 
\label{exQT1}
{\em 
Consider the XQuery query:
``Find all patients
with Admission/Problem = `Coronary' whose Treatment
started Dec, 25th 2003'', expressed as: 
\vspace{-3mm}
\begin{footnotesize} 
\begin{tt}
\begin{tabbing} 
00\=1234\=1234\=1234\=1234\=1234\=1234\=12341234123412341234123412341234 \kill \\ 
\vspace*{-1.5ex} 
\>Q1:
  \>FOR \$A IN //Admission, \\ 
\>\>\>\$P IN //Patient[@ID=\$A/@PatRef] \\ 
\>WHERE \$A/Problem="Coronary" AND \\ 
\>\>\>\$P/Treat/Date="12/25/2003" \\ 
\>RETURN \{\$P/Name\} 
\end{tabbing}
\end{tt} 
\end{footnotesize} 
\vspace{-3mm}


This query can be represented as the join of two tree patterns, 
as shown in Figure~\ref{TP}(a)-(b). In the previous section, we discussed 
how using rule $1$ of Figure~\ref{fig-mapping-rules}, this join query 
can be translated to the tree pattern in Figure~\ref{subst1}(e). 
But that is not yet the translation of query Q1 itself: we have to 
consider translation via other mapping rules in Figure~\ref{fig-mapping-rules} 
as well. A quick inspection reveals rule $2$ is not relevant to the query 
as the rule heads do not contain variables corresponding to the 
variables in the query of  Figure~\ref{TP}. Consider the translation 
via rule $3$. The expanded TP is shown in Figure~\ref{subst2}(a). Its 
corresponding translated query is shown in Figure~\ref{subst2}(b). 
The translation w.r.t. rule $4$ is shown in Figure~\ref{subst2}(c) 
(expanded TP omitted). The contracted translated queries are shown in 
Figures~\ref{subst2}(d). An important point to note is that based on the 
equality between \$ID and \$PR, the two {\tt //Progress} 
nodes (which respectively correspond to $f2(\$ID)$ and $f2(\$PR)$, with $\$ID = \$PR$) 
have been merged. 

Finally, in Figure~\ref{subst2}(e), we combine the translated query 
obtained via rules $1$, $3$, and $4$ (see also Figure~\ref{subst1}(e) 
and Figure~\ref{subst2}(d)). The key point to notice is that since the variables 
\$ID and \$PR have been equated in the mapping rules, we can infer that 
the attributes @ID and @PatRef must be equal. This is again the join of two TPs and 
the corresponding XQuery is given in Figure~\ref{subst2}(f). } 
\end{namedexample}

When $Q$ is a query expressed against the DTD $\Delta_2$, the key 
intuition for query translation is to essentially follow the 
mapping rule in the reverse direction, i.e., from the 
head to the body. This has resemblances to query folding and 
answering queries using views \cite{divesh}. However, the presence of 
Skolem functions greatly simplifies this process. The reason is that the 
node id's act as a ``glue'' suggesting which subelement pieces should be 
associated together. Consequently, they drive exactly which 
mapping rule bodies we need to ``join'' together to rewrite the 
given query. We illustrate this with an example. 

\begin{figure}[htb]
\centering
\includegraphics[height=0.5in,width=2.5in]{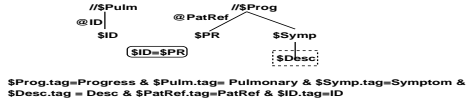}
\vspace{-0.25cm}
\caption{Join of TPs corresponding to Q2.}\label{TP2}
\end{figure}

\begin{figure}[htb]
\begin{tabular}{c} 
\includegraphics[height=0.75in,width=2.0in]{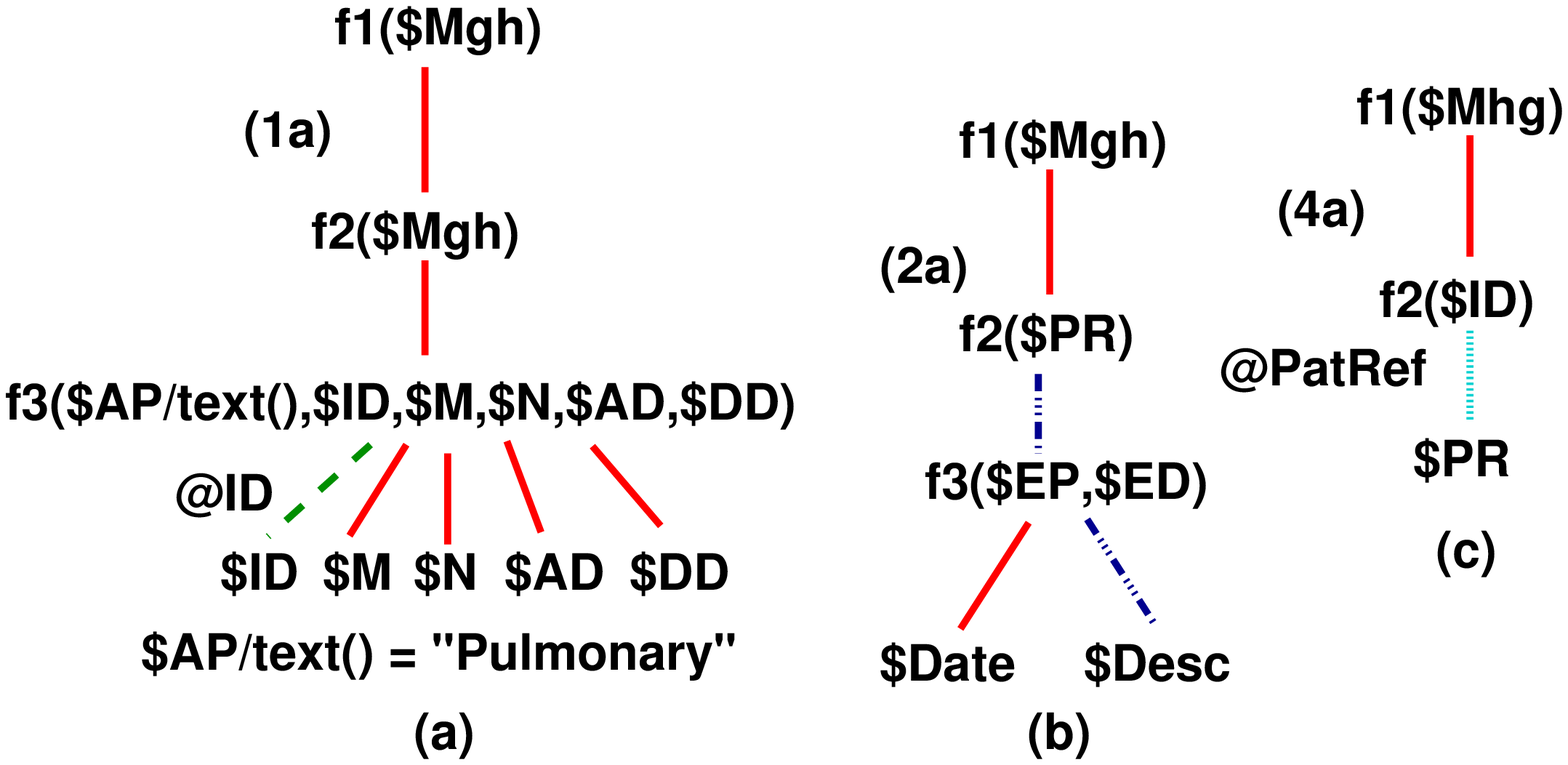} \\ 
\includegraphics[height=0.75in,width=2.5in]{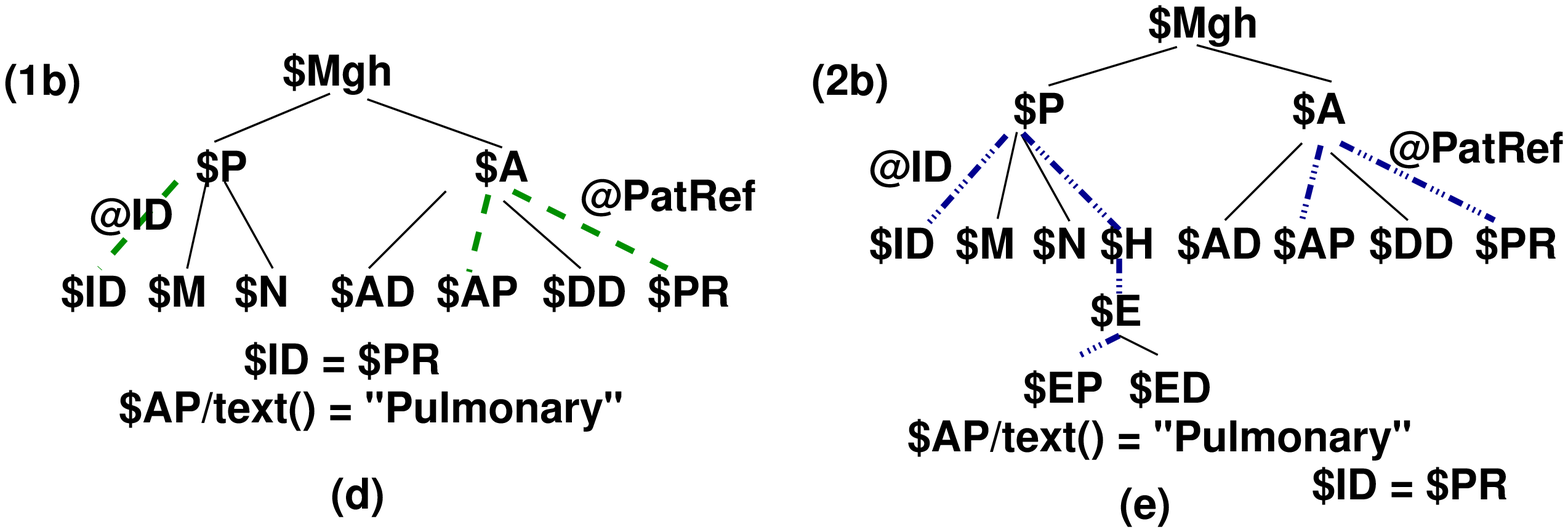} \\ 
\includegraphics[height=0.75in,width=2.5in]{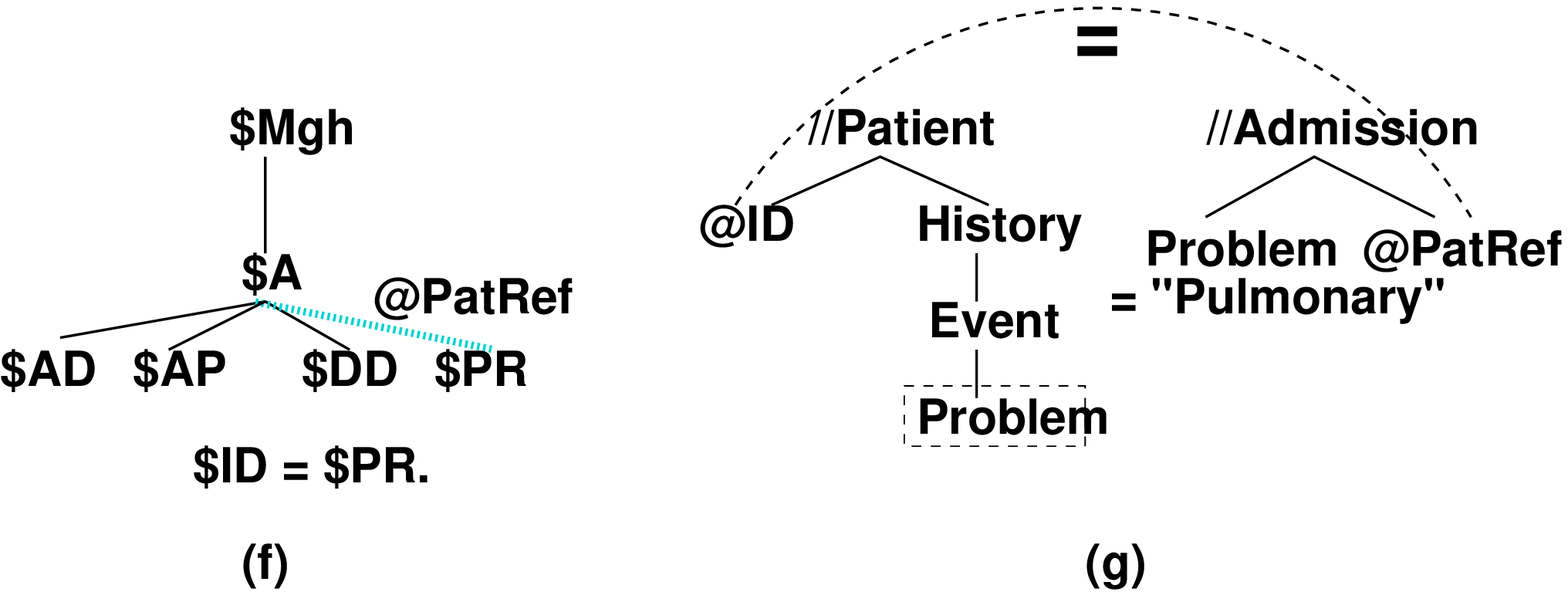} 
\end{tabular} 
\vspace*{-1.5ex} 
\caption{Steps in the Translation of Q2; red = solid, thick; green =
  dashed, thick; blue = dashed \& dotted, thick; cyan = dashed thick.}\label{fused}
\end{figure}

\begin{namedexample}{XQuery Backward} 
\label{exQT2}
{\em 
Consider the XQuery query:
``Find symptoms of patients
admitted with `Pulmonary' (ailment)''. 
\vspace{-3mm}
\begin{footnotesize} 
\begin{tt}
\begin{tabbing} 
00\=1234\=1234\=1234\=1234\=1234\=1234\=12341234123412341234123412341234 \kill \\ 
\vspace*{-1.5ex} 
\>Q2:
  \> FOR \$P IN //Progress, \\ 
\>\>\>    \$Y IN //Pulmonary \\ 
\>WHERE \$Y/@ID=\$P/@PatRef \\
\>RETURN \{\$P/Symptom/Desc\}
\end{tabbing}
\end{tt} 
\end{footnotesize} 
\vspace{-3mm}                                                                                           
Figure~\ref{TP2} shows the join of TPs corresponding to this query. 
The next step is to translate each TP by matching it to each rule 
head. It is easy to see that only the heads of $1, 2, 4$ can be 
matched to the query (partially).\footnote{Actually, in the head 
of rule $3$ just {\tt Progress} can be matched to the query, but 
this is subsumed by matches to other rule heads and is redundant.} 
The expanded TPs are obtained by matching the TPs against each of these 
rule heads and are shown in Figure~\ref{fused}(a)-(c), where to 
minimize clutter, we do not show tag constraints nor the 
substitutions between query variables and the terms in the rule heads. 
In essence, as for the forward direction of translation, each 
expanded TP is a rule head expressed as a tree expression. From the 
names of variables and tags the substitution between query variables and 
rule variables should be implicitly clear. 
We also mark the variables that appeared in the original query (shown 
via distinctly colored edges in the figure). E.g., in 
Figure~\ref{fused}(a)-(c), we know \$Pulm is associated with the node 
$f_3(\$AP/text(), \$ID, \$M, \$N, \$AD, \$DD)$, \$ID with the 
node \$ID, \$Prog with node $f_2(\$PR)$, \$Symp with node $f_3(\$EP, \$ED)$, 
\$Desc with node \$Desc, \$Prog also with the node $f_2(\$ID)$, and finally 
\$PR with node \$PR. 

Next, each expanded TP is replaced by the tree expression in the 
corresponding rule body (Figure~\ref{fused}(d)-(f)). In doing so, we again keep track of 
variables mentioned in the original query. For lack of space, we 
give just two examples of this. Query variable \$Pulm corresponds 
to each of the nodes labeled \$AP in Figure~\ref{fused}(d)-(f) and 
query variable \$Desc corresponds to the node labeled \$EP in 
Figure~\ref{fused}(e). The reader should be able to work out the 
other correspondences with a modest effort. 

The next step is to drop dummy nodes. The idea is very similar to that 
adopted for the forward direction of translation and is not elaborated 
further. Dropping of dummy nodes generates simplified but equivalent TPs. Nodes in different 
TPs that correspond to the same query variable are stitched together. 
E.g., the \$A and \$PR nodes in Figure~\ref{fused}(e) and (f) are merged. 
The final result of merging nodes is shown in Figure~\ref{fused}(g), 
which is actually a join of two TPs. The TPs are shown more concisely 
by writing the tags directly in place of the variables 
that are constrained by those tags. The join of TPs actually 
corresponds to the following XQuery statement: 
\vspace{-4mm}
\begin{footnotesize} 
\begin{tt}
\begin{tabbing} 
00\=1234\=1234\=1234\=1234\=1234\=1234\=12341234123412341234123412341234 \kill \\ 
\vspace*{-1.5ex} 
\>FOR \$P IN //Patient,\\
    \>\>\$A IN //Admission[@ID=\$P/@ID]\\
\>WHERE \$A/Problem='Pulmonary' \\ 
\>RETURN \{\$P/Event\}
\end{tabbing}
\end{tt} 
\end{footnotesize} 
}
\vspace*{-1.5ex} 
\end{namedexample} 

\begin{figure}[ht]
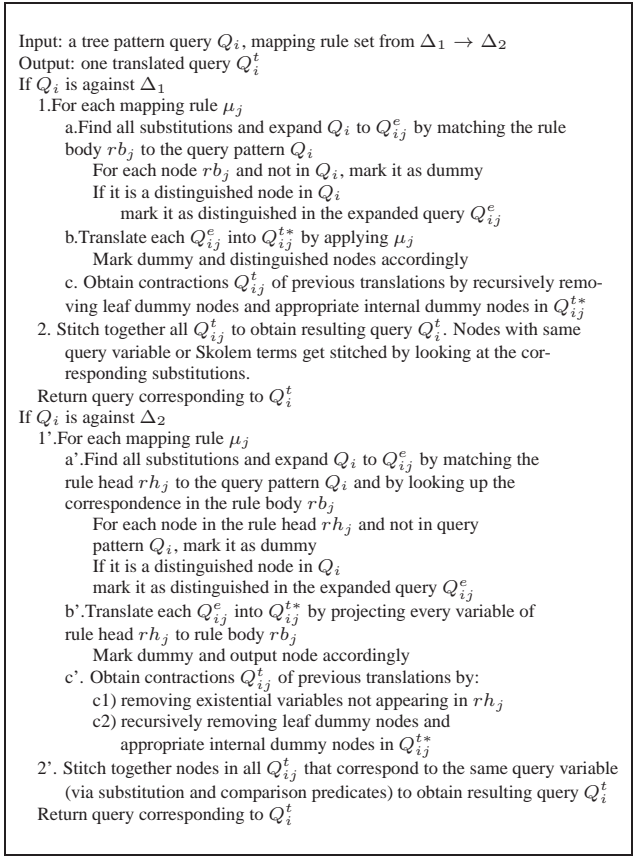

\framebox{
\begin{minipage}{3.25in}
\begin{scriptsize}
\begin{tabbing}
00\=123\=123\=123\=123\=123\=123\=123123123123123123123123 \kill \\
Input: a tree pattern query $Q_i$, mapping rule set from $ \Delta_1 \rightarrow \Delta_2$ \\
Output: one translated query $Q_i^t$ \\
If $Q_i$ is against $\Delta_1$ \\
   \>1.For each mapping rule $\mu_j$\\
   \>\>a.Find all substitutions and expand $Q_i$ to $Q_{ij}^e$ by matching the rule\\
   \>\>  body $rb_j$ to the query pattern $Q_i$\\
   \>\>\> For each node $rb_j$ and not in $Q_i$, mark it as dummy\\
   \>\>\> If it is a distinguished node in $Q_i$\\
   \>\>\>\> mark it as distinguished in the expanded query $Q_{ij}^e$\\
   \>\>b.Translate each $Q_{ij}^e$ into $Q_{ij}^{t*}$ by applying $\mu_j$\\
   \>\>\> Mark dummy and distinguished nodes accordingly\\
   \>\>c. Obtain contractions $Q_{ij}^t$ of previous translations by recursively remo-\\
   \>\>ving leaf dummy nodes and appropriate internal dummy nodes in $Q_{ij}^{t*}$\\
   \>2. Stitch together all $Q_{ij}^t$ to obtain resulting query $Q_i^t$. Nodes with same\\
   \>\> query variable or Skolem terms get stitched by looking at the cor-\\
   \>\> responding substitutions.\\
   \>  Return query corresponding to $Q_i^t$\\
If $Q_i$ is against $\Delta_2$ \\
   \>1'.For each mapping rule $\mu_j$ \\
   \>\>a'.Find all substitutions and expand $Q_i$ to $Q_{ij}^e$ by matching the\\
   \>\>rule head $rh_j$ to the query pattern $Q_i$ and by looking up the\\
   \>\>correspondence in the rule body $rb_j$\\
   \>\>\> For each node in the rule head $rh_j$ and not in query \\
   \>\>\> pattern $Q_i$, mark it as dummy\\
   \>\>\> If it is a distinguished node in $Q_i$\\
   \>\>\> mark it as distinguished in the expanded query $Q_{ij}^e$\\
   \>\>b'.Translate each $Q_{ij}^e$ into $Q_{ij}^{t*}$ by projecting every variable of\\
   \>\>rule head $rh_j$ to rule body $rb_j$\\
   \>\>\> Mark dummy and output node accordingly\\
   \>\>c'. Obtain contractions $Q_{ij}^t$ of previous translations by:\\
   \>\>\>c1) removing existential variables not appearing in $rh_j$\\
   \>\>\>c2) recursively removing leaf dummy nodes and  \\
   \>\>\>\> appropriate internal dummy nodes in $Q_{ij}^{t*}$\\
   \>2'. Stitch together nodes in all $Q_{ij}^t$ that correspond to the same query variable\\
   \>\>(via substitution and comparison predicates) to obtain resulting query $Q_i^t$ \\
   \>  Return query corresponding to $Q_i^t$\\
\end{tabbing}
\end{scriptsize}
\end{minipage}
}
\vspace*{-1.5ex}
\caption{Query Translation Algorithm}
\label{fig-query-alg}
\end{figure}

The algorithm for query translation is given in Figure~\ref{fig-query-alg}, 
and it closely follows the major steps outlined in Examples~\ref{exQT1} and ~\ref{exQT2}. 
Note that the algorithm shown handles one TP at a time. Handling joins of TPs follows the 
same general steps as discussed in the examples. 
We have the following result concerning the correctness of the 
algorithm w.r.t. query answering semantics defined in Definition~\ref{def-sem}. 


\vspace*{-3mm}                  
\begin{namedtheorem}{Correctness of query translation} 
{\em 
For the fragment of XQuery defined in Section~\ref{sec-qta}, 
the query translation algorithm in Figure~\ref{fig-query-alg} is correct 
w.r.t. the semantics of query answering defined in Definition~\ref{def-sem}. \qed } 
\end{namedtheorem}
\vspace*{-3mm}
For space reasons, we omit the proof. The proof is based on an adaptation of the 
well-known chase proof procedure. 
Before leaving this section, we note that our query translation algorithm is 
currently able to handle limited forms of nested XQuery expressions, as 
long as the query can be represented using TPs (with joins). 


\nop{ 
Finally, we discuss here the extensions needed to 
handle GTPs. We remind that the main features of GTPs are the optional
(dashed) edges and the presence of groups.
We can claim the following theorem:

\begin{theorem}
Query answering over GTPs is correct and complete
if Algorithm~\ref{fig-query-alg} is applied to 
each group in the GTP separately and if 
the nodes connected by dashed edges are 
treated as output distinguished nodes.
\end{theorem}

We omit the proof for brevity.
We give instead an example of query answering with
GTPs. Consider the following query:
\begin{verbatim}
Q3:
FOR $P IN //Patient
WHERE $P/MedCr#=0273
RETURN {$P/Name,
        FOR $T IN $P/Treat
        RETURN {$T/desc}}
\end{verbatim}
 
The generated GTP on DTD $\Delta_1$ and
their translations to DTD $\Delta_2$ are 
illustrated in Figure~\ref{fig-gtp}.

\begin{figure}[htb]
\centering
\includegraphics[height=2.0in,width=2.5in]{gtp1.eps}
\caption{Example with GTPs for Query Q3.}\label{fig-gtp}
\end{figure}

We invite the reader to perform for each group
the steps indicated by Algorithm~\ref{fig-query-alg}.
The final translated query will look like the 
following:
\begin{verbatim}
FOR $Pr IN //Admission/*
WHERE $Pr/Policy#=0273
RETURN {$Pr/Patient,
        FOR $P IN //Progress,
            $T IN $P/Treatment
        WHERE $Pr/ID=$P/PatRef 
        RETURN {$T/desc}}
\end{verbatim}
}


\vspace*{-1ex} 
\section{Experiments} 
\label{sec-exper} 

\noindent {\bf Implementation and Setup}: 
In this section, we examine how the query 
translation algorithm impacts the query performance and the scalability of 
HepTox.
Using the mapping rules discussed in Section~\ref{sec-arrow2rules}, each peer maps to 
a set of acquaintances.  Transitive mappings produces semantic paths, and 
each peer is subsequently semantically connected to every other node. By means of a 
preliminary experiment, we realized that in order to 
obtain a reasonably small number of hops in network with number of peers $<=$ 128, 
we have to consider sets of acquaintances having at least 4 peers. 
Henceforth, we assume to have such sets of 
acquaintances for all experiments, which guarantee an 
average number of hops equal to 5.

Our aim is to observe how our approach behaves in a network of XML database 
systems having heterogeneous schemas. Hence, we derived 9 \emph{restructured 
variations} of the XMark DTD~\cite{xmark}, corresponding to the various 
transformations expressible in our mapping rule language, to produce 10 different 
schemas randomly scattered across the network.  Detailed description of the 
schemas can be found on the HepToX home page~\cite{heptoxHP}. 
We modified the XMark xmlgen code accordingly to generate the data sets
 corresponding to the schemas above, with sizes ranging $10MB$-$100MB$.
We use Emulab~\cite{emulab}, a network emulation testbed, to emulate a realistic
P2P database. Emulab consists of a collection of PCs, whose network
delay and bandwidth can be set at wish. We chose a 70ms delay and a
 50MB bandwidth to simulate as much as possible the geographical networks behavior.
We could get $30$ real machines for emulation.
As network protocol among emulab machines, we mounted 
FreePastry~\cite{pastry}, which offers a scalable and efficient 
P2P routing algorithm. 
Its O(logN) routing complexity, and O(logN) routing table size is at 
least as efficient as other similar projects such as CAN, Tapestry and Chord. 
Heptox integrates FreePastry with QIZX~\cite{qizx}, which is 
apparently the fastest available open-source XQuery engine. 
We used FreePastry vs.1.3.2 and QIZX vs.0.4p1, respectively.

\noindent 
{\bf Guiding Principles}: 
The experiments are conducted according to the following guidelines: 
\emph{(i)} peers gets created within the Pastry network, 
each peer owning an independent dataset, stored in its own QIZX query engine;
\emph{(ii)} each peer selects other peers as its acquaintances, and these 
acquaintances in turn see this peer as their acquaintance;
\emph{(iii)} then a peer initiates a query, translates the query 
to the schema of each acquaintance, using our algorithm, 
and sends the translated query to that acquaintance;
\emph{(iv)} these acquaintances in turn translate the query and send it to 
their acquaintances, which process the query and return all the results backward 
to the originating peer;
\emph{(v)} forwarding of queries by a peer stops as soon as the peer realizes it 
received an already processed query request.

\noindent 
{\bf Scalability of HepToX}:
The first experiment is devoted to measuring the efficiency of the Heptox query 
translation algorithm (Figure~\ref{fig-query-alg}). 
\vspace*{-2mm}
\begin{table}[htb!]~\label{tab1}
\begin{center}
\scriptsize{
\begin{tabular}{|c|c||c|c|c|}
\hline
{\em Q} & {\em Query Descr.} & {\em AQTT} & {\em MQTT} & {\em ASPL}\\
\hline\hline
$Q_1$ & TP with only '/' & 1180.02 & 2714 & 2.38 \\
$Q_2$ & TP with 1sel & 749.06   & 1581 & 2.93 \\
$Q_3$ & TP with 1sel,1join & 2057.57    & 4480 & 2.97 \\
$Q_4$ & TP with 2sel,1join & 2236.26    & 4566 & 2.4 \\
$Q_5$ & TP with 2sel,1join & 1749.56    & 3637 & 2.63\\
$Q_6$ & TP with 2join & 2054.27 & 4246 & 2.98\\
$Q_7$ & TP with 1sel,2join & 2299.3     & 4651 & 2.42\\
$Q_8$ & TP with 1sel,1join & 1379.53    & 3013 & 2.41\\
$Q_9$ & TP with 2sel,2join & 2472.03    & 4996 & 2.87\\
$Q_{10}$ & TP with 2sel,3join & 3172.41 & 7069 &  2.92 \\
\hline
\end{tabular}
\caption{AQTT (Average Query Translation Time in $ms$); MQTT (Maximum Query Translation Time in $ms$);
ASPL (Average Semantic Path Length).}\vspace*{-1.5ex}
}
\end{center}
\end{table}
\vspace{-2mm}

\begin{figure}[t]
\begin{center} 
\includegraphics[width=3in, height=1.6in,origin=2in]{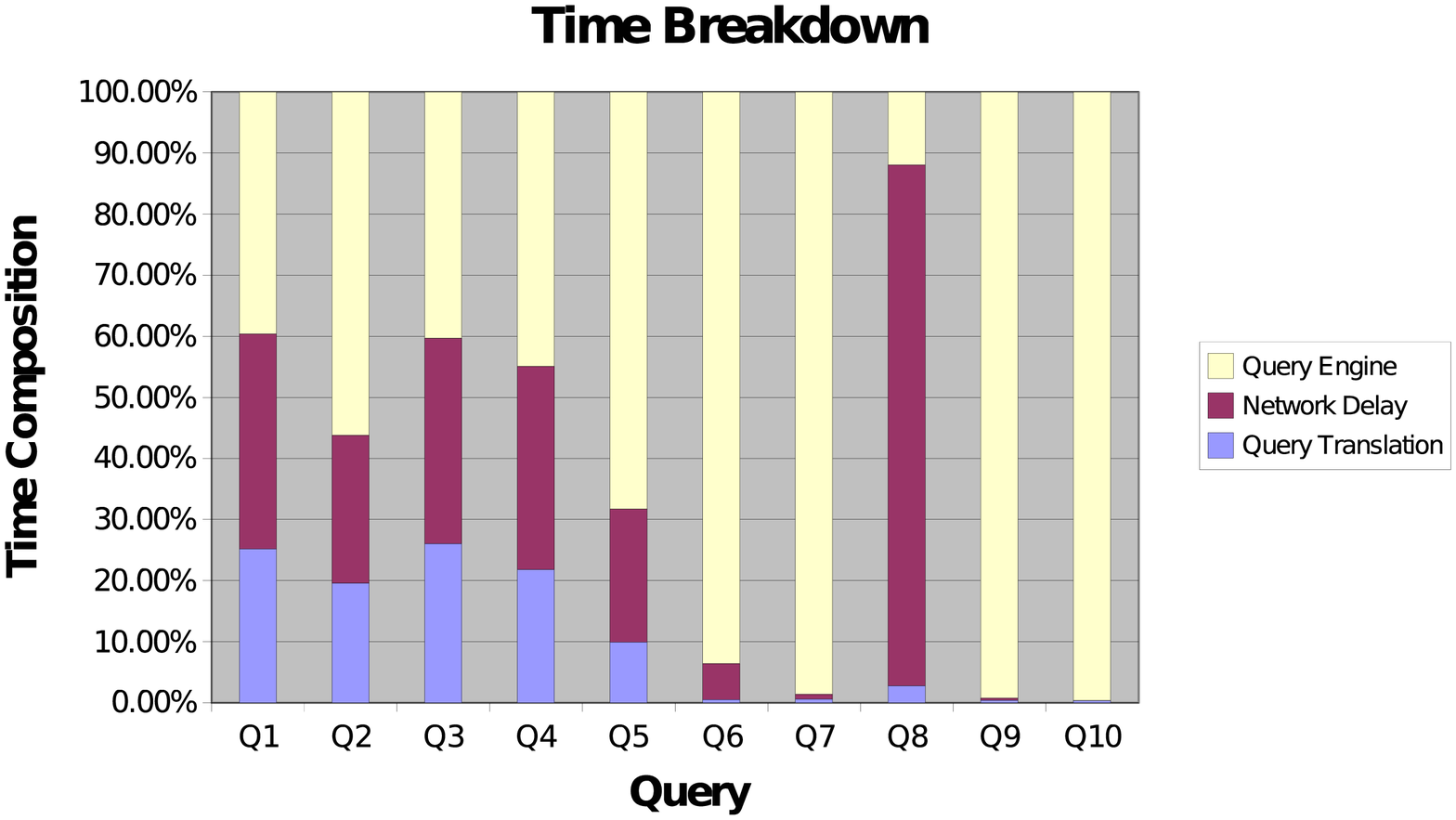} 
\includegraphics[width=3in,height=1.6in]{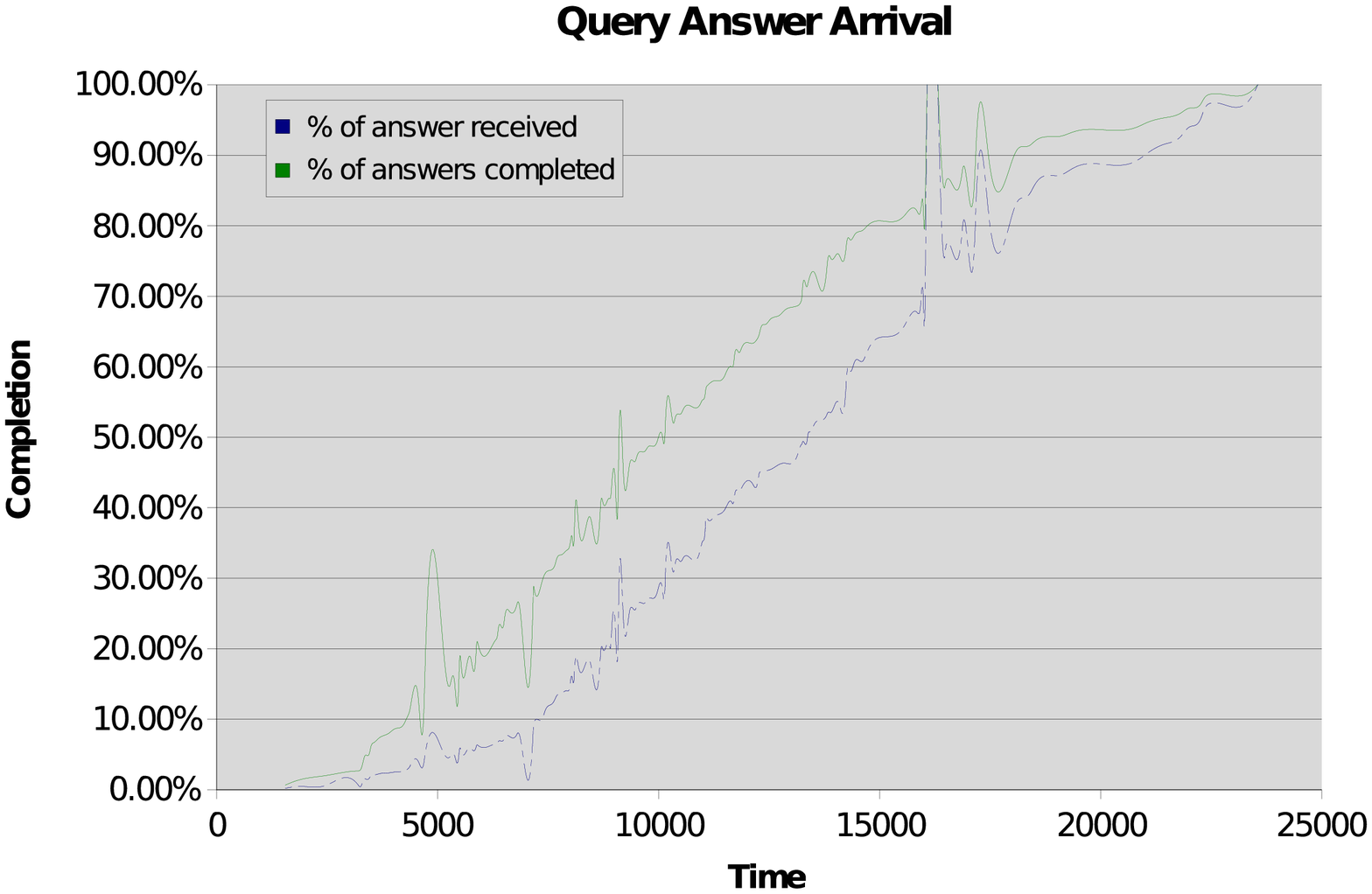} 
\end{center}
\vspace{-7mm}
\caption{Query Answering Time Breakdown (top) 
and Time distribution of Answer Arrival at Source (bottom).
\label{fig:exp}} 
\end{figure}

We derived 10 queries of increasing 
complexity in both structure and constraints (see brief description in 
Table 1, while complete queries can be 
be found at~\cite{heptoxHP}).  
We broadcasted the above queries 
from one peer to the rest of the network (across all the acquaintances), 
and measured the average and maximum time 
taken to translate each query along all semantic paths. Table 1 shows
that the translation times stay within few seconds, and increasing 
query complexity (e.g. query $Q_{10}$) slightly increases the translation time.
%
%
In the second experiment, we aim at showing the impact of schemas heterogeneity
on HepTox performance. Therefore, we vary the number of 
distinct heterogeneous schemas scattered around the network, and 
measure the average query translation time along all the semantic paths.  As 
expected, Table 2 
shows the query translation time goes from $2$ (two heterogeneous 
schemas across the network), to $10$ (ten heterogeneous schemas).  
Table 2 shows that the translation time grows linearly with
the number of schemas considered and that it also depends on 
the complexity of the translated query. For instance, queries $Q_6, Q_7, Q_9$ and
$Q_{10}$ having double joins and multiple selections, have an higher translation 
time.
%
\begin{table}[htb!]~\label{tab2}
\begin{center}
\scriptsize{
\begin{tabular}{|c||c|c|c|c|c|c|c|c|c|} 
\hline
 &\multicolumn{9}{c|}{\em Nr.of Dist.Schemas}\\
{\em Q} & {\em 2} & {\em 3} & {\em 4} & {\em 5} & {\em 6} & {\em 7} & {\em 8} & {\em 9} & {\em 10}\\
\hline\hline
$Q_1$  & 193 & 397 &	595 & 779 & 998 & 1142 & 1133 & 1194 &
1274 \\
$Q_2$  & 126 & 231 &	377 & 505 & 680 & 844 & 834 & 872 & 944 \\
$Q_3$  & 249 & 548 &	899 & 1187 & 1543 & 1788 & 1808 & 1909 & 2075 \\
$Q_4$  & 276 & 576 & 947 & 1261 & 1629 & 1907 & 1945 & 2036 & 2155 \\
$Q_5$  & 211 & 460 & 761 & 1027 & 1331 & 1557 & 1570 & 1651 & 1773 \\
$Q_6$  & 327 & 642 & 1064 &	1381 & 1795 & 2074 & 2096 &	2209	 & 2328 \\
$Q_7$  & 333 & 658 &	1108 & 1437	& 1869 & 2189 & 2209 &	2324 &
2441 \\
$Q_8$  & 212 & 430  & 705 & 911 & 1186 & 1376 & 1495 & 1471 & 1593 \\
$Q_9$  & 308 & 657 &	1078	 & 1440 & 1855 & 2179 & 2215 & 2340& 2562 \\
$Q_{10}$  &	546 & 973 & 1513 & 1977 & 2545 & 2975 & 2948 & 3170 & 3275 \\
\hline
\end{tabular}
}
\caption{Query Translation times ($ms$) with varying number of distinct schemas (obviously
equal to $0$ with $1$ schema).}\vspace*{-1.5ex}
\end{center}
\end{table}
%
The next experiment wants to show the minimal overhead introduced by our 
translation algorithm all 
along the query answering process.  In Figure~\ref{fig:exp}(top), we measure the 
average time taken by each query to be completed. 
We highlight the various time components taken by query translation, 
network delay, and local query answering, respectively.  
It can be noted that query translation takes a negligible time
 if compared to network delay and local 
query answering. Local query answering, essentially 
imputable to QIZX, was the bottleneck
for all queries and caused the crashing of the most complex ones. 
Query answers to $Q_7$, $Q_9$ and $Q_{10}$ show a cutoff point 
due to the fact that they never completed and were timed out.
We tried different query engines before choosing QIZX, which is
actually the fastest, thus this behavior was somehow outside our control.


\noindent 
{\bf Query Performances in HepToX}: 
To probe the impact of our query translation algorithm on the system's performance, 
we conducted an experiment in the Emulab environment in which a query is issued at an 
originating peer and spread across the entire network. We measured 
the percentage of query answers reaching the originating 
peer at each time interval, along with the percentage of 
peers completed.
Figure~\ref{fig:exp}(bottom) shows the results as a cumulative plot. The answers arrive 
in regular bursts just as expected in any normal P2P system. 


\nop{ 
It can be noted that, although the distribution has some peaks, 
the overall trend of $\%$ of answers completed (top curve)
closely tracks the $\%$ of answers received (bottom curve).  
} 

Finally, we conducted several experiments on simultaneous query processing, which is 
also feasible in HepTox. From these experiments, we realized that,
similarly to what happens when one query is issued, 
the main component of time with multiple issued queries, 
is caused by local query answering, while
the translation time still stays small. For lack of space, we omit
these results.






\vspace*{-1ex} 
\section{Related Work} 
\label{sec-rw} 

{\bf P2P data integration systems}
We discuss here Piazza~\cite{med03, wwwhalevy03}, Hyperion~\cite{miller-sigrec},
Clio~\cite{millerClio03} and Lenzerini's et al.~\cite{lenzP2P}. 
Piazza~\cite{med03, wwwhalevy03} defines semantic LAV/GAV-style mappings
among peer schemas. Query answering semantics is
based on the notion of certain answers~\cite{med03}, and is based on 
query answering using views \cite{divesh}, a problem not yet  
completely solved for XML. Our mappings are easier and more intuitive to specify,  
and have a data exchange semantics (similar to~\cite{miller-sigrec}) 
rather than a view semantics.
Query answering also follows the same principle.
In Piazza, mappings need to be provided in an XQuery-like language, 
while in HepTox we can infer them from diagrams.
Hyperion~\cite{miller-sigrec} combines mapping expressions and mapping
tables.
Mapping tables being value-based correspondences across the world, 
are orthogonal to mapping expressions. Mapping tables 
provide a lightweight mechanism for sharing data.
Our Datalog-like mapping rules follow the same direction 
but capture schema correspondences rather than data value ones. 
Clio~\cite{millerClio03, popa04} also deals with 
XML data integration by considering 
semantically independent target and source
nested schemas enriched with nested referential constraints. 
We found that this is very similar in 
spirit with what HepTox does.
However, ~\cite{millerClio03, popa04, med03, wwwhalevy03} 
are limited in the heterogeneity they deal with and
do not capture the class of XML transformations
involving data schema interplay (see Figure~\ref{fig-running-eg}).
Lenzerini et al.~\cite{lenzP2P} address the problem of
data interoperation in P2P systems using expressive schema mappings, also 
following the GAV/LAV paradigm, and 
show that the 
problem is in PTIME only when mapping rules are
expressed in epistemic logic. 

{\bf Schema-matching systems}
Automated techniques for
schema matching (e.g. CUPID ~\cite{bernstein01}, \cite{rahm01, potti03}) 
are able to
output elementary schema-level associations by exploiting
linguistic features, context-dependent type matching, similarity 
functions etc. These associations 
could constitute the input of our rule inference algorithm
if the user does not provide the arrows. 

{\bf P2P systems with non-conventional lookups}
Popular P2P networks, e.g. Kazaa, Gnutella, advertise simple lookup queries on 
file names. The idea of building full-fledged P2P DBMS
is being considered in many works.
Internet-scale database queries and functionalities~\cite{PierDB}
as well as approximate range queries in P2P~\cite{elAbbadi} and 
XPath queries in small communities of peers~\cite{dewitt03} 
have been extensively dealt with.
All these works do not deal with reconciling schema heterogeneity.
~\cite{dewitt03} relies on 
a DHT-based network to address simple XPath queries, while~\cite{peerDB-ng-etal-icde03}
realizes IR-style queries in an efficient P2P relational database.




\vspace*{-1ex} 
\section{Conclusions and Future Work} 
\label{sec-summ} 

We have presented the HepToX P2P XML database system, focusing on the 
following key conceptual contributions: (i) an algorithm for 
inferring mapping rules from correspondences between heterogeneous 
peer DTDs, specified via boxes and 
arrows; (ii) a precise and intuitive semantics for query evaluation in 
a P2P setting; (iii) a query translation algorithm that is correct w.r.t. 
this semantics and is efficient, as revealed by the detailed experimentation. 
We are currently investigating larger fragments of XQuery. A fundamental 
challenge is the reconstruction of the answers obtained from query translation in 
the schema of the originating peer.
Another important milestone is handling 
1-n and m-n correspondences between elements across DTDs in the sense of 
\cite{rahm01}. A promising recent work in this direction is 
\cite{doan+sigmod04}. It would be interesting to capture such complex mappings within 
our framework. 

%
\begin{small}
\bibliographystyle{plain}
\bibliography{heptox-vldb}  
\end{small}
%
%
\end{document}